%% file: paper.tex
\documentclass[12pt]{iopart}
\usepackage{subfiles}
\usepackage{amsfonts}
\expandafter\let\csname equation*\endcsname\relax
\expandafter\let\csname endequation*\endcsname\relax 
\usepackage{amsmath}
\usepackage{amssymb}
\usepackage{hyperref}
\usepackage{mathtools}
\usepackage{xcolor}

\begin{document}

\title[A Dean-Kawasaki equation for reaction diffusion systems driven by Poisson noise]{A Dean-Kawasaki equation for reaction diffusion systems driven by Poisson noise}

\author{Richard E. Spinney}
\address{School  of  Physics, University  of  New  South  Wales  -  Sydney  2052,  Australia}
\address{EMBL-Australia  node  in  Single  Molecule  Science, School of Medical Sciences, University  of  New  South  Wales  -  Sydney  2052,  Australia}
\author{Richard G. Morris}
\address{School  of  Physics, University  of  New  South  Wales  -  Sydney  2052,  Australia}
\address{EMBL-Australia  node  in  Single  Molecule  Science, School of Medical Sciences, University  of  New  South  Wales  -  Sydney  2052,  Australia}
\address{ARC Centre of Excellence for the Mathematical Analysis of Cellular Systems‌, UNSW Node, Sydney, NSW 2052, Australia}

\ead{r.spinney@unsw.edu.au}
\vspace{10pt}

\begin{indented}  
	\item[]\today
\end{indented} 

\begin{abstract}
	We derive a stochastic partial differential equation that describes the fluctuating behaviour of reaction-diffusion systems of $N$ particles, undergoing Markovian, unary reactions. This generalises the work of Dean [J. Phys. A: Math. and Gen., 29 (24), L613, (1996)] through the inclusion of random Poisson fields. Our approach is based on weak interactions, which has the dual benefit that the resulting equations asymptotically converge (in the $N\to\infty$ limit) on a variation of a McKean-Vlasov diffusion, whilst still being related to the case of Dean-like strong interactions via a trivial rescaling. Various examples are presented, alongside a discussion of possible extensions to more complicated reaction schemes.
\end{abstract}

\subfile{content}

\ack{The authors thank J. Worsfold for useful discussions and close reading of this manuscript. RES and RGM acknowledge funding from the EMBL Australia program. RGM acknowledges funding from the Australian Research Council Centre of Excellence for Mathematical Analysis of Cellular Systems (CE230100001).‌}  

\section*{References}


\subfile{references}

\end{document}

%% file: content.tex

\section{Introduction}
\label{model}

The Dean-Kawasaki (DK) equation \cite{deanLangevinEquationDensity1996,kawasakiStochasticModelSlow1994,kawasakiMicroscopicAnalysesDynamical1998} is a fundamental equation of fluctuating hydrodynamics which, despite certain mathematical challenges \cite{konarovskyiDeanKawasakiDynamics2020, konarovskyiDeanKawasakiDynamicsIllposedness2019,cornalbaRegularizedDeanKawasaki2019,cornalbaWellposednessRegularisedInertial2021}, concisely represents the fluctuating evolution of the empirical density of many body systems of interacting particles. 
Whilst its origins can be traced back to the work of Kawasaki, who described the evolution of a coarse grained probability density functional \cite{kawasakiStochasticModelSlow1994,illienDeanKawasakiEquationStochastic2024}, the DK equation is synonymous with the explicitly fluctuating dynamic introduced later by Dean \cite{deanLangevinEquationDensity1996}. 
Such an equation, for identical unit diffusions interacting through a pair potential $U$, takes the form
\begin{align}
    \label{DKeqn}
\dot{\rho}(x,t)&=\nabla_{\!x}\cdot\left(\rho(x,t)\int dy\,\rho(y,t)\nabla_{\!x} U(x-y)\right)+\frac{1}{2}\nabla_{\!x}^2\rho(x,t)-\nabla_{\!x}\cdot\left(\sqrt{\rho(x,t)}\xi(x,t)\right).
\end{align}
The hallmark of this equation resides in its final, fluctuating, term, containing a vector valued Gaussian white noise, $\xi(x,t)$. Paired with the multiplying square root density, such a term ensures that the statistics of a test function integrated against the empirical field, $\rho$, are equivalent to those found from the underlying particles. 
\par
This equation, however, only concerns interacting diffusions of a conserved number of a single species of particle. In this contribution we generalise the DK equation (Eq.~(\ref{DKeqn})) to include Markovian unary reaction kinetics, characteristic of, for example, biological switching \cite{bressloffStochasticSwitchingBiology2017}. This corresponds to the facilitation of transitions between species that are characterised by Poisson noise, whose rates are coupled to the density fields. The motivation and result mirrors those of the original contribution due to Dean \cite{deanLangevinEquationDensity1996} --- we seek a concise fluctuating dynamic for the collective density field(s), formulated without approximation. To do so we describe a stochastic partial differential equation (SPDE) that is driven, in part, by random spatio-temporal \emph{Poisson} fields. The resulting equation is similar to one offered in \cite{bressloffGlobalDensityEquations2024}, which concerns switching particles, but does so at the level of the mean in the switching behaviour, and is thus absent such noise terms. 
We anticipate that such equations, being statistically exact at all fluctuation scales, may prove useful in the theory and analysis of systems where one is concerned about systems with low particle number or large atypical fluctuations, where Poisson noise substantially deviates from a limiting Gaussian form. Such situations arise in the hydrodynamic treatment of mesoscale biological agents \cite{julicherHydrodynamicTheoryActive2018, marchettiHydrodynamicsSoftActive2013} and broadly in macroscopic fluctuation theory \cite{bertiniMacroscopicFluctuationTheory2015}/large deviation theory \cite{touchette_large_2009}.
\par
In a departure from Dean's original contribution \cite{deanLangevinEquationDensity1996}, our approach broadly follows that offered in \cite{worsfoldDensityFluctuationsStochastic2023}, which presents an SPDE describing the empirical \emph{probability} density field of \emph{weakly coupled} particles, which asymptotically converges on a McKean-Vlasov diffusion \cite{mckeanClassMarkovProcesses1966,jabinMeanFieldLimit2017}, through a propagation of chaos result \cite{chaintronPropagationChaosReview2022, chaintronPropagationChaosReview2022a}. This has the particular advantage that a limiting (particle number taken to infinity) microscopic dynamic can be associated with a closed deterministic evolution equation for the probability density. We note, however, that conversion between the weakly coupled description underpinning McKean-Vlasov diffusions and the perhaps more familiar strong coupling found in Dean is trivial, and so such a conversion is offered in a section where these differences are explored.
\par
The outline of the paper is as follows. First, we describe the microscopic dynamics that we wish to represent as a fluctuating density dynamic, and establish a vector space for particles of different species to reside in. Next, we describe the dynamics of individual particles in terms of couplings to the vector field of particle species using inner products on the vector space. This then allows for a description of the evolution of a test function on the fields establishing a weak form of a PDE, up to stochastic terms which cannot be written without reference to the stochastic behaviour of the individual particles. This allows for a brief discussion of the dynamic of the mean behaviour, its relation to the propagation of chaos results, and the notion of weak coupling. We then present a jump SPDE, based on spatio-temporal Poisson processes and Gaussian white noise. This is shown to possess the same statistics for the empirical average of a test function, as the underlying particulate system. Notably, unlike in the usual formulation of the DK equation, this requires matching of all jump moments in the empirical average, not just the demand for equality in the two-point correlation function. Using this equation we present two simple examples that illustrate its application to a system with constant, and field dependent, species interconversion rates, respectively. We also discuss its Gaussian limiting form as the number of particles is taken to infinity. Finally, before concluding, we discuss the possibility of future application to reaction diffusion systems formed of higher order reactions (binary, ternary \emph{etc.}) and those, generally, where particle number is not conserved.
\section{Microscopic description}
We consider a system of $N$, identical, weakly coupled point particles with continuous state $ X_i(t)\equiv(X_i^{(1)}(t),\ldots,X_i^{(d)}(t))\in \Omega$, $i\in\{1,\ldots,N\}$, $t\in\mathbb{R}$, with $\Omega$ a connected subset of $\mathbb{R}^d$ with suitable boundary conditions, and discrete internal state $\mathcal{Y}_i(t)\in\mathcal{A}$, $|\mathcal{A}|=m\geq 1$, with piece-wise constant, \emph{c\'adl\'ag}, sampling paths. The continuous $X_i$ evolve in time according to first order stochastic differential equations (SDEs)
\begin{align}
    X^{(k)}_i(t) &=X^{(k)}_i(0)+ \frac{1}{N}\sum_{j=1}^N (1-\delta_{i,j}) \int_0^t\mathcal{K}^{(k)}_{\mathcal{Y}_i(t'),\mathcal{Y}_j(t')}(X_i(t'),X_j(t'))\mathop{dt'}\nonumber\\
    &\quad\qquad+\sqrt{2D}\int_0^t\mathop{dW^{(k)}_i}(t'),\quad\forall\, i\in\{1,\ldots,N\},\, k\in\{1,\ldots,d\}. 
    \label{SDEi}
\end{align}
The functions $\mathcal{K}^{(k)}_{\cdot,\cdot}(\cdot,\cdot):\mathcal{A}^2\times\Omega^2\to \mathbb{R}$ are spatial components of a $\mathcal{Y}_i,\mathcal{Y}_j$-pair dependent interaction kernel, allowing the deterministic drift of a particle to depend on the positions and internal states of the other particles (in addition to its own), $D$ is an isotropic diffusion constant, and the $dW^{(k)}_i(t)$ are increments of independent Wiener processes such that $\mathbb{E}[dW^{(k)}_i(t)dW^{(l)}_j(s)]=\delta_{i,j}\delta_{k,l}\delta(t-s)dt$, 
with $\delta_{\cdot,\cdot}$ the Kronecker delta, $\delta(\cdot)$ the Dirac delta, and $\mathbb{E}[\cdot]$ an expectation. 
Weak coupling is implemented here through the $N^{-1}$ pre-factor to the interaction kernel, with self interaction explicitly excluded. All stochastic integrals are interpreted in the It\^{o} sense, here and throughout. Where convenient we will express the above set of equations in terms of vectors in $\mathbb{R}^d$
\begin{align}
    X_i(t) &=X_i(0)+ \frac{1}{N}\sum_{j=1}^N (1-\delta_{i,j}) \int_0^t \mathcal{K}_{\mathcal{Y}_i(t'),\mathcal{Y}_j(t')}(X_i(t'),X_j(t'))\mathop{dt'}\nonumber\\
    &\qquad\qquad\qquad+\sqrt{2D}\int_0^t\mathop{dW_i}(t'),\quad\forall\, i\in\{1,\ldots,N\},
    \label{SDEv}
\end{align}
where $dW_i(t)\equiv(dW_i^{(1)}(t),\ldots,dW_i^{(d)}(t))$ and $\mathcal{K}_{\cdot,\cdot}(\cdot,\cdot):\mathcal{A}^2\times\Omega^2\to \mathbb{R}^d$, such that $\mathcal{K}_{\cdot,\cdot}(\cdot,\cdot)\equiv(\mathcal{K}^{(1)}_{\cdot,\cdot}(\cdot,\cdot),\ldots,\mathcal{K}^{(d)}_{\cdot,\cdot}(\cdot,\cdot))$.
\par
Note, if we take the case $|\mathcal{A}|=m=1$ and $\mathcal{K}_{\mathcal{Y}_i(t),\mathcal{Y}_i(t)}(X_i,X_j)=-N\nabla_{\!X_i}U(X_i-X_j)$, where $\nabla_{\!X_i}=(\partial_{X_i^{(1)}},\ldots,\partial_{X_i^{(d)}})$ and $U$ is a pair potential function, we recover the original dynamic considered by Dean \cite{deanLangevinEquationDensity1996}. 
\par
The internal states, $\mathcal{Y}_i(t)$, evolve through memory-less Poissonian jumps with intensity of a transition for a given particle dependent on the positions and states of all other particles (in addition to its own). Since the internal states, $\alpha\in\mathcal{A}$, need not be numeric, we keep track of them by considering  a vector space, $V_{\mathcal{A}}$, on $\mathbb{R}$ with basis formed from states $\mathcal{A}$ such that we have basis vectors $|\alpha\rangle\in V_{\mathcal{A}}$, corresponding co-vectors $\langle \alpha|\in V_{\mathcal{A}}^*$, and inner product $\langle \alpha|\beta\rangle=\delta_{\alpha,\beta}$, $\alpha,\beta\in\mathcal{A}$. We can therefore write the dynamics of the internal state (vector) for an individual particle, with probability one,
\begin{align}
    |\mathcal{Y}_i(t)\rangle &=|\mathcal{Y}_i(0)\rangle+\int_0^t\sum_{\alpha\in\mathcal{A}\setminus\{\mathcal{Y}_i(t')\}}\Big[|\alpha\rangle-|\mathcal{Y}_i(t')\rangle\Big]\mathop{dN^{(i)}_{{\mathcal{Y}_i,\alpha}}}(t'),\quad\forall\, i\in\{1,\ldots,N\},
    \label{dYSDE}
\end{align}
where $dN^{(i)}_{{\alpha,\beta}}(t)=\lim_{dt\to 0}N^{(i)}_{{\alpha,\beta}}(t)-N^{(i)}_{{\alpha,\beta}}(t-dt)\in\{0,1\}$ is an increment in one of $N\,m\,(m-1)$  inhomogeneous Poisson processes, $N^{(i)}_{{\alpha,\beta}}(t)$, (one for each possible transition/directed pair in $\mathcal{A}$, for each particle). 
All such processes are independent in the sense that at some time, $t$, we have, almost surely, $dN^{(i)}_{{\alpha,\beta}}(t)dN^{(j)}_{\gamma,\delta}(t)=dN^{(i)}_{{\alpha,\beta}}(t)\delta_{i,j}\delta_{\alpha,\gamma}\delta_{\beta,\delta}$. 
\par
Further, we may collect all such Poisson processes for an individual particle into the linear operator, $d\mathbf{N}^{(i)}_t$, allowing Eq.~(\ref{dYSDE}) to be written, equivalently, as
\begin{subequations}
    \begin{align}
    |{\mathcal{Y}}_i(t)\rangle &=|{\mathcal{Y}}_i(0)\rangle+\int_0^t d\mathbf{N}_{t'}^{(i)}\,|\mathcal{Y}_i(t')\rangle,\quad\forall i\in\{1,\ldots,N\},\\
    \langle \beta|d\mathbf{N}_t^{(i)}|\alpha \rangle &=\begin{cases}
    dN^{(i)}_{\alpha,\beta}(t),&\beta\neq \alpha,\\
    -\sum_{\gamma\in\mathcal{A}\setminus\{\alpha\}}dN^{(i)}_{\alpha,\gamma}(t),&\beta=\alpha.
    \end{cases}
    \label{dNdef}
    \end{align}
\end{subequations}
The intensity of each individual Poisson process, $dN_{\alpha,\beta}^{(i)}(t)$, which we write as $h_{\alpha,\beta}^{(i)}(t)$, is thus equal to the rate at which particle $i$ makes the transition $\alpha\to\beta$. This rate depends on the state of all other particles, $\{X(t),\mathcal{Y}(t)\}\equiv\{\{X_1(t),\mathcal{Y}_1(t)\},\ldots,\{X_N(t),\mathcal{Y}_N(t)\}\}$, through the kernel function $\mathcal{H}^{\alpha,\beta}_{\cdot,\cdot}(\cdot,\cdot):\mathcal{A}^2\times\Omega^2\to \mathbb{R}$, and mediating function, $H_{\alpha,\beta}(\cdot):\mathbb{R}\to\mathbb{R}$, which allows for eventual non-linear dependencies on the resulting field, \emph{viz.} 
\begin{align}
    h^{(i)}_{\alpha,\beta}(t)\equiv h^{(i)}_{\alpha,\beta}(\{X(t),\mathcal{Y}(t)\})&=\delta_{\alpha,\mathcal{Y}_i(t)}H_{\alpha,\beta}\left[\frac{1}{N}\sum_{j=1}^N(1-\delta_{i,j}) \mathcal{H}^{\alpha,\beta}_{\mathcal{Y}_i(t),\mathcal{Y}_j(t)}(X_i(t),X_j(t))\right].
    \label{particlerate}
\end{align}
The individual $dN^{(i)}_{{\alpha,\beta}}(t)$ take value $1$ at  the time of a transition, and $0$ otherwise, such that we may identify $\mathbb{E}[dN_{\alpha,\beta}^{(i)}(t)|\{X(t),\mathcal{Y}(t)\}]=h^{(i)}_{\alpha,\beta}(t)dt$. 

\par
We may then write the probability mass function for the internal state of particle $i$ as the vector $|P_i\rangle=\sum_{\alpha\in\mathcal{A}}\mathcal{P}_{i,\alpha}(t)|{\alpha}\rangle$, where $\langle\alpha|P_i\rangle=\mathcal{P}_{i,\alpha}(t)$ is the probability that the internal state of particle $i$ is in state $\alpha$, at time $t$.  The probability mass vectors associated with each particle then obey the non-autonomous master equations
\begin{subequations}
    \begin{align}
    |\dot{P}_i\rangle &=\mathbf{H}_t^{(i)}\,|P_i\rangle,\quad\forall i\in\{1,\ldots,N\},\\
    \langle \beta|\mathbf{H}_t^{(i)}|\alpha \rangle 
    &=\begin{cases}
    h^{(i)}_{\alpha,\beta}(t),&\beta\neq \alpha,\\
    -\sum_{\gamma\in\mathcal{A}\setminus\{\alpha\}}h^{(i)}_{\alpha,\gamma}(t),&\beta=\alpha,
    \end{cases}
    \end{align}
\end{subequations}
with linear operator $\mathbf{H}_t^{(i)}$
acting as the generator for the internal state of particle $i$.
\par
Finally, we specify the initial state of the particles, at time $t=0$, to be independent, and drawn from some distribution, $p_0^\alpha(x)$, satisfying $\sum_{\alpha\in\mathcal{A}}\int_\Omega dx\; p_0^\alpha(x)=1$, $p_0^\alpha(x)\geq 0$, and remark that the choices of interaction kernels, $\mathcal{K}_{\cdot,\cdot}(\cdot,\cdot)$, $\mathcal{H}_{\cdot,\cdot}^{\cdot,\cdot}(\cdot,\cdot)$ and mediation functions $H_{\cdot,\cdot}(\cdot)$ are assumed to be restricted to cases such that the above microscopic SDEs are well-posed.

\section{Dynamics dependent on empirical density vectors}
\label{DynDenVec}
On the vector space, $V_{\mathcal{A}}$, we then consider an empirical (\emph{i.e.} stochastic, or fluctuating) one-particle probability density field (or random measure), $|\phi\rangle$, comprised of one-body empirical fields for each internal state $|\phi^\alpha\rangle=\phi^\alpha(x,t)|\alpha\rangle$, $x\in\Omega$, $\alpha\in\mathcal{A}$. This total empirical probability density vector,  $|\phi\rangle$, is then defined simply as $|\phi\rangle \coloneq  \sum_{\alpha\in\mathcal{A}} |\phi^\alpha\rangle$. We may also consider the marginal empirical scalar field $\phi(x,t)=\langle\mathbf{1}|\phi\rangle=\sum_{\alpha\in\mathcal{A}}\phi^\alpha(x,t)$, where $\langle\mathbf{1}|\coloneq \sum_{\alpha\in\mathcal{A}}\langle\alpha|$. Each particle is represented as a Dirac measure, which, for narrative convenience, we write as delta (generalised) functions   
$|\phi_i\rangle =\delta(x-X_i(t))|\mathcal{Y}_i(t)\rangle=\prod_{j=1}^d\delta(x^{(j)}-X^{(j)}_i(t))|\mathcal{Y}_i(t)\rangle$, 
such that we can write $\int_\Omega dx\;\delta(x-X_i(t))=1$. 
We then write the one-body empirical probability density vectors $|\phi^\alpha\rangle=\frac{1}{N}\sum_{i=1}^N|\alpha\rangle\langle\alpha|\phi_i\rangle$ such that 
$|\phi\rangle =\sum_{\alpha\in\mathcal{A}}|\phi^\alpha\rangle=\frac{1}{N}\sum_{i=1}^N|\phi_i\rangle$. 
\par
If we consider a set of smooth test functions with compact support, $F_\alpha(\cdot):\Omega\to\mathbb{R}$, we can construct the co-vector $\langle F|\coloneq \sum_{\alpha\in\mathcal{A}}\langle \alpha|F_\alpha(x)$, mapping vectors in $V_\mathcal{A}$ to the space of distributions on $\Omega$, $\langle F|\,\cdot\,\rangle:V_{\mathcal{A}}\to\mathcal{D}(\Omega)$, and define the (time-dependent) spatial inner product with the field, $\langle F,\phi\rangle_t$, $\langle F,\,\cdot\,\rangle_t:V_{\mathcal{A}}\to\mathbb{R}$, as
\begin{align}
\langle F,\phi\rangle_t &\coloneq \int_{\Omega} dx\;\langle F|\phi\rangle =
\sum_{\alpha\in\mathcal{A}}\int_\Omega dx\;F_\alpha(x)\phi^\alpha(x,t).
\end{align}
The (vector) field $|\phi\rangle$ then fulfils the role of an empirical probability density by the normalisation property $\langle\mathbf{1},\phi\rangle_t=1$. Note, where clear from context, we may omit the subscript indicating the time dependence of such inner products.
\par
In cases of vector functions in $\mathbb{R}^d$, such as $\mathcal{K}_{\cdot,\cdot}(\cdot,\cdot)$, or some other set of functions $\mathcal{F}_\alpha(\cdot):\Omega\to\mathbb{R}^d$, we understand that the inner product of two such vector quantities is also performed with respect to $\mathbb{R}^d$, such that co-vectors, $\langle \mathcal{F}|$, are implicitly also co-vectors in $\mathbb{R}^d$, \emph{i.e.} $\langle\mathcal{F}|\coloneq \sum_{\alpha\in\mathcal{A}}\langle\alpha|\mathcal{F}_\alpha^T(x)$, with $\cdot^T$ a transpose operation. We thus have, for example,
\begin{align}
    \langle \mathcal{F},\nabla\phi\rangle_t\equiv \sum_{\alpha\in\mathcal{A}}\int_\Omega dx \;\mathcal{F}^T_\alpha(x) \nabla_{\!x}\phi^\alpha(x,t)\equiv \sum_{\alpha\in\mathcal{A}}\int_\Omega dx \;\mathcal{F}_\alpha(x) \cdot \nabla_{\!x}\phi^\alpha(x,t),
\end{align}
as an appropriate scalar quantity, where $\nabla_{\!x}\equiv (\partial_{x^{(1)}},\ldots,\partial_{x^{(d)}})$.
\par
Next, due to the atomic nature of the fields, we can then write population weighted summations of a function of individual particle states, $f^{(1)}_\cdot(\cdot):\mathcal{A}\times\Omega\to\mathbb{R}$, as inner products of the field with the co-vector $\langle f^{(1)}|\equiv\langle f^{(1)}|(x)\coloneq \sum_{\alpha\in\mathcal{A}}\langle \alpha|f_\alpha^{(1)}(x)$, \emph{viz.}
\begin{align}
\frac{1}{N}\sum_{i=1}^Nf^{(1)}_{\mathcal{Y}_i(t)}(X_i(t))&
 =\frac{1}{N}\sum_{i=1}^N\sum_{\alpha\in\mathcal{A}}\int_\Omega dx\;f^{(1)}_\alpha(x) \delta(x-X_i(t))\langle\alpha|\mathcal{Y}_i(t)\rangle\nonumber\\
 &=\frac{1}{N}\sum_{i=1}^N\sum_{\alpha\in\mathcal{A}}\int_\Omega dx\;\langle f^{(1)}|\alpha\rangle\langle\alpha|\phi_i\rangle
 =\int_{\Omega}dx\;\langle f^{(1)}|\phi\rangle=\langle f^{(1)},\phi\rangle_t.
\end{align}
Similarly, with an abuse of notation, a population weighted summation of a function of $n$ particles, for instance $n=2$ through the function $f^{(2)}_{\cdot,\cdot}(\cdot,\cdot):\mathcal{A}^2\times\Omega^{2}\to\mathbb{R}$, and associated  co-vector $\langle f^{(2)}_{\alpha}|\equiv \langle f^{(2)}_{\alpha}|(x,y) \coloneq \sum_{\beta\in\mathcal{A}}\langle\beta |f^{(2)}_{\alpha,\beta}(x,y)$, we can exchange the summated variables with an inner product,  but leaving a pointwise dependence
\begin{align}
    \frac{1}{N}\sum_{j=1}^Nf^{(2)}_{\mathcal{Y}_i(t),\mathcal{Y}_j(t)}(X_i(t),X_j(t)) &=\frac{1}{N}\sum_{j=1}^N\sum_{\alpha\in\mathcal{A}}\int_\Omega dx\;f^{(2)}_{\mathcal{Y}_i(t),\alpha}(X_i(t),x) \delta(x-X_j(t))\langle\alpha|\mathcal{Y}_j(t)\rangle\nonumber\\
&=\frac{1}{N}\sum_{j=1}^N\sum_{\alpha\in\mathcal{A}}\int_\Omega dx\;\langle\alpha|\phi_j\rangle\langle f^{(2)}_{\mathcal{Y}_i(t)}|\alpha\rangle(X_i(t))\nonumber\\
&=\int_{\Omega}dx\;\langle f^{(2)}_{\mathcal{Y}_i(t)}|\phi\rangle(X_i)=\langle f^{(2)}_{\mathcal{Y}_i(t)},\phi\rangle(X_i(t)).
\end{align}
In the particular case  $f^{(2)}_{\mathcal{Y}_i(t),\mathcal{Y}_j(t)}(X_i(t),X_j(t))= f_{\mathcal{Y}_i(t)}(X_i(t)-X_j(t))$, relevant to applications with a local density interaction without dependence on the species of other particles, as explored in the later examples, this inner product is identifiable as a convolution of the form $\langle f_{\mathcal{Y}_i(t)}, \phi\rangle(X_i(t)) \equiv (f_{\mathcal{Y}_i(t)}*\phi)(X_i(t))\coloneq \sum_{\alpha\in\mathcal{A}} \int_\Omega dx\;f_{\mathcal{Y}_i(t)}(X_i(t)-x)\phi^\alpha(x,t)=\int_\Omega dx\;f_{\mathcal{Y}_i(t)}(X_i(t)-x)\phi(x,t)$.
\par
Consequently, with such sums represented as inner products with the field, the single particle dynamics, dependent on the state of all other particles, can be written, up to $\mathcal{O}(1/N)$ correction terms (to account for the self interaction exclusions $N^{-1}\mathcal{K}_{\mathcal{Y}_i(t),\mathcal{Y}_i(t)}(X_i(t),X_i(t))$ and $N^{-1}\mathcal{H}^{\alpha,\beta}_{\mathcal{Y}_i(t),\mathcal{Y}_i(t)}(X_i(t),X_i(t))$), in terms of the single particle state coupled to the field
\begin{subequations}
    \begin{align}
    X_i(t) &=X_i(0) +\int_0^t\langle \mathcal{K}_{\mathcal{Y}_i(t')}, \phi\rangle(X_i(t'))\mathop{dt'}+ \sqrt{2D}\int_0^t\mathop{dW_i(t')},\\
|\mathcal{Y}_i(t)\rangle&=|\mathcal{Y}_i(0)\rangle+\int_0^t\mathop{d\mathbf{N}^{(i)}_{t'}}|\mathcal{Y}_i(t')\rangle,\\
h^{(i)}_{\alpha,\beta}(t)&=\delta_{\alpha,\mathcal{Y}_i(t)}H_{\alpha,\beta}\left[\langle\mathcal{H}^{\alpha,\beta}_{\mathcal{Y}_i(t)},\phi\rangle(X_i(t))\right],
\label{particletofield}
\end{align}
\end{subequations}
for all $i\in\{1,\ldots,N\}$, 
where $\langle\mathcal{K}_{\mathcal{Y}_i(t)}, \phi\rangle(X_i(t))\coloneq (\langle\mathcal{K}^{(1)}_{\mathcal{Y}_i(t)}, \phi\rangle(X_i(t)),\ldots,\langle\mathcal{K}^{(d)}_{\mathcal{Y}_i(t)}, \phi\rangle(X_i(t)))$, \emph{i.e.} shorthand notation for a spatial vector in $\mathbb{R}^d$ of inner products on $V_\mathcal{A}$.
\par
We note the approximation arising from the self interaction exclusion at this stage can be elided entirely by utilising kernels that satisfy $\mathcal{K}^{(i)}_{\alpha,\beta}(X_i(t),X_i(t))=0$ and $\mathcal{H}^{\alpha,\beta}_{\alpha,\gamma}(X_i(t),X_i(t))=0$. In this case the interaction summations, of the type appearing in Eqs.~(\ref{SDEi}) and (\ref{particlerate}), can be performed without the self-interaction exclusion.
\section{Evolution of the empirical average of a test function}
With coupling to other particles expressed as an interaction with the empirical field, $|\phi\rangle$, we consider the evolution of the inner product of the field and above test co-vector $\langle F|$, $\langle F,\phi\rangle_t$, representing the empirical average of $\langle F|$. 
Since $|\phi\rangle$ is merely a linear sum of the $|\phi_i\rangle$, we first compute the contribution for each $|\phi_i\rangle$, before combining them, leveraging the property that $|\phi_i\rangle$ varies only through changes in $X_i$ and $\mathcal{Y}_i$. Next, because we have diffusions and jumps in the underlying dynamics we can describe a change in the contribution $\langle F,\phi_i\rangle_t$, with the appropriate generalised version of It\^{o}'s lemma.
In our case we have  purely discontinuous changes in $\mathcal{Y}_i(t)$ and purely continuous changes in $X_i(t)$ allowing us to write the change in $\langle F,\phi_i\rangle_t$ 
as
\begin{align}
    \langle F,\phi_i\rangle_t-\langle F,\phi_i\rangle_0
        &=\int_0^t(\nabla_{\!X_i} F_{\mathcal{Y}_i(t')}(X_i(t'))) \cdot \langle \mathcal{K}_{\mathcal{Y}_i(t')},\phi\rangle(X_i(t'))\mathop{dt'}\nonumber\\
        &\quad+\sqrt{2D}\int_0^t(\nabla_{\!X_i}  F_{\mathcal{Y}_i(t')}(X_i(t')))\cdot \mathop{dW_i(t')}+ D\int_0^t\nabla^2_{\!X_i} F_{\mathcal{Y}_i(t')}(X_i(t')) \mathop{dt'}\nonumber\\
&\quad+\int_0^t\sum_{\beta\in\mathcal{A}\setminus\{\mathcal{Y}_i(t)\}}(F_\beta(X_i(t'))-F_{\mathcal{Y}_i(t')}(X_i(t')))h^{(i)}_{\alpha,\beta}(t')\mathop{dt'}\nonumber\\
&\qquad+\int_0^t\sum_{\beta\in\mathcal{A}\setminus \{\mathcal{Y}_i(t')\}}(F_\beta(X_i(t'))-F_{\mathcal{Y}_i(t')}(X_i(t')))\mathop{d\hat{N}^{(i)}_{{\mathcal{Y}_i(t'),\beta}}(t')},
\label{dFi}
\end{align}
with $\nabla^2=\nabla\cdot\nabla$, and where we have introduced the compensated Poisson processes, $\hat{N}^{(i)}_{{\alpha,\beta}}(t)$, with increments $d\hat{N}^{(i)}_{{\alpha,\beta}}(t)\coloneq dN^{(i)}_{{\alpha,\beta}}(t)-h^{(i)}_{\alpha,\beta}(t) dt$, and key property $\mathbb{E}[d\hat{N}^{(i)}_{{\alpha,\beta}}(t)]=0$. 
Collecting the contributions from all particles (recalling $N^{-1}\sum_{i=1}^N|\phi_i\rangle=|\phi\rangle$) then allows us to express the total change in the empirical average as
\begin{align}
   \langle F,\phi\rangle_t-\langle F,\phi\rangle_0&=N^{-1}\sum_{i=1}^N\left[\langle F,\phi_i\rangle_{t}-\langle F,\phi_i\rangle_{0}\right]\nonumber\\
   &=\int_0^t\left[\langle \nabla F,\mathbf{K}_{\phi}\phi\rangle_{t'} +D\langle \nabla^2 F,\phi\rangle_{t'} +\langle F,\mathbf{H}_\phi\,\phi\rangle_{t'}\right] \mathop{dt'} \nonumber\\
   &\qquad+N^{-1}\sum_{i=1}^N\int_0^t\left\{\sqrt{2D}\langle \nabla F,\phi_i\,dW_i(t')\rangle +\langle F,d\hat{\mathbf{{N}}}_{t'}^{(i)}\,\phi_i\rangle\right\},
   \label{dY}
\end{align}
where $\langle \nabla F|\coloneq \sum_{\alpha\in\mathcal{A}}\langle\alpha|\nabla_{\!x} F_\alpha(x)$, we use notation  $\langle \psi,\mathbf{A}\phi\rangle\equiv\langle \psi\mathbf{A},\phi\rangle\equiv\int_{\Omega}dx\;\langle \psi|\mathbf{A}|\phi\rangle$, and each compensated Poisson linear operator, $d\hat{\mathbf{{N}}}_t^{(i)}:=d{\mathbf{{N}}}_t^{(i)}-\mathbf{H}_t^{(i)}\mathop{dt}$, is constructed from its uncompensated version, $d\mathbf{N}_t^{(i)}$ (Eq.~(\ref{dNdef})). We have then defined the non-linear operator (in $\phi^\alpha$), $\mathbf{H}_\phi$, acting on the total field vector $|\phi\rangle$, through elements 
\begin{subequations}
\begin{align}
\langle\beta|\mathbf{H}_\phi|\alpha\rangle&=\begin{cases}
    h_{\alpha,\beta}(x,t),&\alpha\neq\beta,\\
    -\sum_{\gamma\in\mathcal{A}\setminus\{\alpha\}} h_{\alpha,\gamma}(x,t),&\alpha=\beta,
\end{cases}\\
h_{\alpha,\beta}(x,t)&=H_{\alpha,\beta}\left[\langle\mathcal{H}^{\alpha,\beta}_{\alpha},\phi\rangle(x,t)\right],\\
\langle\mathcal{H}^{\alpha,\beta}_{\alpha},\phi\rangle(x,t)&=\sum_{\gamma\in\mathcal{A}}\int_\Omega \mathcal{H}_{\alpha,\gamma}^{\alpha,\beta}(x,y)\phi^\gamma(y,t)\,dy.
\end{align}
\end{subequations}
Similarly, the deterministic component of the continuous dynamics is described in terms of the non-linear (in $\phi^\alpha(x,t)$) diagonal operator on the field, $\langle\beta|\mathbf{K}_{\phi}|\alpha\rangle=\langle \mathcal{K}_\alpha,\phi\rangle(x,t)\delta_{\alpha,\beta}$, with
\begin{align}
    \langle\mathcal{K}_\alpha,\phi\rangle(x,t)&=\sum_{\gamma\in\mathcal{A}}\int_\Omega \mathcal{K}_{\alpha,\gamma}(x,y)\phi^\gamma(y,t)\,dy.
\end{align}
Integrating by parts over $\Omega$  then gives a description of the evolution of the inner product, solely in terms of other inner products over the test function
\begin{align}
    \langle F,\phi\rangle_t-\langle F,\phi\rangle_0&=\int_0^t\left[-\langle F,\nabla\cdot(\mathbf{K}_{\phi}\,\phi)\rangle_{t'} + D\,\langle F,\nabla^2\phi\rangle_{t'} +\langle F,\mathbf{H}_\phi\,\phi\rangle_{t'} \right]\mathop{dt'} \nonumber\\
    &\qquad-N^{-1}\sum_{i=1}^N\int_0^t\left\{\sqrt{2D}\langle F,\nabla \cdot(\phi_i\mathop{dW_i(t')})\rangle -\langle F,d\hat{\mathbf{{N}}}^{(i)}_{t'}\,\phi_i\rangle\right\}.
\end{align}
However, we are prevented from considering this as a weak form of a field dynamic since the stochastic components cannot be written in terms of the total probability density vector, $|\phi\rangle$.
\section{BBGKY hierarchy and propagation of chaos limit as a single particle dynamic}
 We can proceed without the problematic noise terms by taking suitable expectations. If we take ensemble averages over the evolution and initial conditions of the empirical field, denoted  $\mathbb{E}[\cdot]$, of the above SDE in the $t\to 0$ limit, terms with increments of Wiener and compensated Poisson noise vanish, since they are Martingales, leaving, after exchanging integration order through Fubini's theorem,
\begin{align}
\langle F,\partial_t\mathbb{E}[\phi]\rangle&=-\langle F,\nabla\cdot\mathbb{E}[\mathbf{K}_{\phi}\,\phi]\rangle +D\,\langle F,\nabla^2 \mathbb{E}[\phi]\rangle +\langle F,\mathbb{E}[\mathbf{H}_\phi\,\phi]\rangle, 
\label{ensembleF}
\end{align}
since $d\langle F,\phi\rangle_t/dt = \langle F,\partial_t{\phi}\rangle$, stressing that $\phi$ is the fluctuating variable - not the $F_\alpha$ or $x$. Since this holds for all $\langle F|$, we may write the implied integro-PDE as
\begin{align}
    \partial_t \mathbb{E}[\phi^\alpha(x,t)]&=\sum_{\beta\in\mathcal{A}\setminus\{\alpha\}}\mathbb{E}\left[H_{\beta,\alpha}\left[\sum_{\gamma\in\mathcal{A}}\int_\Omega dy\;\mathcal{H}^{\beta,\alpha}_{\beta,\gamma}(x,y)\phi^{\gamma}(y,t)\right]\phi^{\beta}(x,t)\right]\nonumber\\
    &\qquad-\sum_{\beta\in\mathcal{A}\setminus\{\alpha\}}\mathbb{E}\left[H_{\alpha,\beta}\left[\sum_{\gamma\in\mathcal{A}}\int_\Omega dy\;\mathcal{H}^{\alpha,\beta}_{\alpha,\gamma}(x,y)\phi^{\gamma}(y,t)\right]\phi^{\alpha}(x,t)\right]\nonumber\\
    &\qquad+\sum_{\gamma\in\mathcal{A}}\int_\Omega dy\; \left[-\nabla_{\!x}\cdot \mathcal{K}_{\alpha,\gamma}(x,y)+D\nabla^2_{\!x} \right]\mathbb{E}[{\phi}^{\alpha}(x,t){\phi}^{\gamma}(y,t)],\quad\forall\alpha\in\mathcal{A},
\end{align}
leaving a non-closed forward equation for the mean, $\mathbb{E}[\phi^\alpha(x,t)]$. This is a statement of the BBGKY-hierarchy in this context \cite{cercignaniManyParticleDynamicsKinetic1997}, with closure failing due to the non-linearity arising from the field dependence introduced from the $\mathcal{K}$ and $\mathcal{H}$ interaction kernels. Explicitly, the dynamic for the one-body probability density $p^\alpha(x,t)\equiv\mathbb{E}[\phi^\alpha(x,t)]$ depends on the two-body probability density $p^{\alpha,\beta}(x,y,t)\equiv\mathbb{E}[\phi^\alpha(x,t)\phi^\beta(y,t)]$ (concerning a particle in joint state $\{x,\alpha\}$ and another in $\{y,\beta\}$), along side others depending on the precise form of the mediating functions, $H_{\alpha,\beta}$. In the case of all mediating functions simply being the identity, $H_{\alpha,\beta}(z)=z$,  we may write the forward equations for the 1-body probability densities on $\mathcal{A}\times \Omega$ as explicitly dependent on the 2-body probability densities on $\mathcal{A}^2\times \Omega^2$, \emph{viz.}
\begin{align}
    \partial_t {p}^\alpha(x,t)&=\sum_{\beta\in\mathcal{A}\setminus\{\alpha\}}\sum_{\gamma\in\mathcal{A}}\int_\Omega dy\;\left[\mathcal{H}^{\beta,\alpha}_{\beta,\gamma}(x,y)p^{\beta,\gamma}(x,y,t)-\mathcal{H}^{\alpha,\beta}_{\alpha,\gamma}(x,y)p^{\alpha,\gamma}(x,y,t)\right]\nonumber\\
    &\qquad+\sum_{\gamma\in\mathcal{A}}\int_\Omega dy\; \left[-\nabla_{\!x}\cdot \mathcal{K}_{\alpha,\gamma}(x,y)+D\nabla^2_{\!x} \right]{p}^{\alpha,\gamma}(x,y,t),\quad\forall\alpha\in\mathcal{A}.
\end{align}
However, the so-called propagation of chaos results \cite{chaintronPropagationChaosReview2022, chaintronPropagationChaosReview2022a, sznitmanTopicsPropagationChaos1991} then allow us to close all such dependencies as $N\to\infty$, by factoring joint probability density functions with independent terms such that, for example, ${p}^{\alpha,\beta}(x,y,t)$ terms become ${p}^{\alpha}(x,t){p}^{\beta}(y,t)$, effecting a deterministic mean field description through the equivalent limit $\mathbb{E}[\phi^\alpha(x,t)\phi^\beta(y,t)]\to\mathbb{E}[\phi^\alpha(x,t)]\mathbb{E}[\phi^\beta(y,t)]$. Note that in this limit, the system's domain, $\Omega$, is \emph{fixed}, whilst the order of magnitude of the sum of all interactions is kept constant such that the interaction between any two individual particles tends to zero in accordance with the prescription of weak coupling.
\par
We also note that a corollary of such vanishing pairwise interactions, in such a limit, is that the $N$-body system becomes statistically equivalent to $N$ identical and \emph{independent} processes, made explicit by recognising that the total probability density has no correlative structure, \emph{i.e.} $p^{\alpha_1,\ldots,\alpha_N}(x_1,\ldots,x_N)=\prod_{i=1}^Np^{\alpha_i}(x_i)$. As such, we may consider the particles in the system no longer coupled to all other particles, but rather coupled to their own probability distribution. Thus, when the limit is taken, the hierarchy collapses into a single, non-linear, integro-master-Fokker-Planck equation, with non-linearity provided through the $\mathcal{K}_{\cdot,\cdot}(\cdot,\cdot)$ and $\mathcal{H}_{\cdot,\cdot}^{\cdot,\cdot}(\cdot,\cdot)$ kernels, which equivalently (in distribution) describes a \emph{single particle}, \emph{viz.}
\begin{align}
    \partial_t {p}^\alpha(x,t)\nonumber&= \sum_{\beta\in\mathcal{A}\setminus\{\alpha\}}H_{\beta,\alpha}\left[\sum_{\gamma\in\mathcal{A}}\int_\Omega dy\;\mathcal{H}^{\beta,\alpha}_{\beta,\gamma}(x,y)p^\gamma(y,t)\right]p^\beta(x,t)\nonumber\\
    &\quad-\sum_{\beta\in\mathcal{A}\setminus\{\alpha\}}H_{\alpha,\beta}\left[\sum_{\gamma\in\mathcal{A}}\int_\Omega dy\;\mathcal{H}^{\alpha,\beta}_{\alpha,\gamma}(x,y)p^\gamma(y,t)\right]p^\alpha(x,t)\nonumber\\
    &\quad-\sum_{\gamma\in\mathcal{A}}\int_\Omega dy\;p^\gamma(y,t)\nabla_{\!x}\cdot \mathcal{K}_{\alpha,\gamma}(x,y) {p}^{\alpha}(x,t)+D\nabla^2_{\!x} {p}^{\alpha}(x,t),\quad\forall\alpha\in\mathcal{A},
    \label{FPMS}
    \end{align}
with associated single particle dynamics
\begin{subequations}
\begin{align}
    \tilde{X}(t) &= \tilde{X}(0)+\sum_{\alpha\in\mathcal{A}}\int_0^tdt'\int_\Omega dx\;\mathcal{K}_{\tilde{\mathcal{Y}}(t'),\alpha}(\tilde{X}(t'),x)p^\alpha(x,t')+ \sqrt{2D}\int_0^t\mathop{d\tilde{W}(t')},\\
     |\tilde{\mathcal{Y}}(t)\rangle&=|\tilde{\mathcal{Y}}(0)\rangle+\int_0^td\tilde{\mathbf{N}}_{t'}|\tilde{\mathcal{Y}}(t')\rangle,\\
     \tilde{h}_{\alpha,\beta}(t)&=\delta_{\alpha,\tilde{\mathcal{Y}}(t)}H_{\alpha,\beta}\left[\sum_{\gamma\in\mathcal{A}}\int_\Omega dx\;\mathcal{H}^{\alpha\beta}_{\alpha,\gamma}(\tilde{X}(t),x)p^\alpha(x,t)\right],
     \label{singleparticle}
\end{align}
\end{subequations}
in terms of a single, vector valued, Wiener process, $d\tilde{W}$, and $m\, (m-1)$ independent Poisson processes, $d\tilde{N}_{\alpha,\beta}$, with intensities $\tilde{h}_{\alpha\beta}(t)$, comprising linear operator $d\tilde{\mathbf{N}}_t$. 
\section{Alternative noise representation and closed evolution equation}
It is perhaps not obvious that such a deterministic limit should exist, and it is desirable to retain a fluctuating description without the average over the noise performed in the previous section. We can heuristically both confirm such a result and retain information about fluctuations in our many-body system by considering a parallel system, which replaces the problematic atomic Wiener and  Poisson noises with carefully chosen random fields, which agree in law for an arbitrary test variable. This allows us to write a closed SPDE in terms of a single probability density vector $|\Phi\rangle$ without reference to individual particles or increments. To do so we consider the evolution of the test variable $Y(t)=\langle F,\phi\rangle_t$, representing the empirical average of $\langle F|$. 
We then construct a proposed dynamics of such an empirical average using the random fields and demand that it has identically distributed increments, by demonstrating that they have equal moments up to $\mathcal{O}(dt)$. For purely continuous underlying SDEs we would achieve this statistical equivalence through a demand for equality in the mean and for equal quadratic variation, $[Y]_t$, in expectation, with no higher moments contributing owing to the Gaussian nature of the Brownian motions on all length scales. 
However, since we are using jump processes with finite intensity, we are obliged to generalise the quadratic variation and consider the $n$-adic variation in expectation, which we define as
\begin{align}
    \mathbb{E}[[Y]^{(n)}_t]&\coloneq \mathbb{E}\left[\int_{0}^{t}(dY(t'))^n\right]=\mathbb{E}\left[\int_{0}^{t}(d\langle F,\phi\rangle_{t'})^n\right].
\end{align}
Given the ``box calculus" properties of It\^o integrals and their generalisation to include jumps, we can write, with probability one, for all $n,n'\in\{i\in\mathbb{Z}\;|\;i > 0\}$, for an arbitrary well-behaved function $f(\cdot):\Omega\to\mathbb{R}$ satisfying $\int_0^tdt'\;f^2(x(t'))<\infty$,
\begin{subequations}
    \begin{align}
        &\int_{0}^{t} f(x(t'))\,dW^{(k)}_i(t')dW^{(l)}_j(t')=\delta_{i,j}\delta_{k,l}\int_0^tf(x(t'))\,dt',\\
        &\int_{0}^{t} f(x(t'))\,d\hat{N}^{(i_1)}_{\alpha_1,\beta_1}(t')\prod_{j=2}^n d\hat{N}^{(i_j)}_{\alpha_j,\beta_j}(t')=\left(\prod_{j=1}^{n-1} \delta_{i_j,i_{j{+}1}}\delta_{\alpha_j,\alpha_{j{+}1}}\delta_{\beta_j,\beta_{j{+}1}}\right)\nonumber\\
            &\qquad\qquad\qquad\qquad\qquad\qquad\qquad\qquad\qquad\times\int_{0}^{t} f(x(t'))\,dN^{(i_1)}_{\alpha_{i_1},\beta_{i_1}}(t'), 
    \end{align}
    and
    \begin{align}
    \int_{0}^{t} f(x(t'))\,dW^{(l)}_i(t')dW^{(m)}_j(t')\prod_{k=1}^n dW^{(d_k)}_k(t')&=0,\\
        \int_{0}^{t}f(x(t'))\, \prod_{i=1}^n dt'\prod_{j=1}^{n'}dW^{(k)}_j(t')&=0,\\
        \int_{0}^{t} f(x(t'))\,\prod_{i=1}^n dt'\prod_{j=1}^{n'}d\hat{N}_{\alpha_j,\beta_j}^{(i_j)}(t')&=0,\\
        \int_{0}^{t} f(x(t'))\,\prod_{i=1}^{n}dW^{(k)}_i(t')\prod_{j=1}^{n'} d\hat{N}^{i_j}_{\alpha_j,\beta_j}(t')&=0.
    \end{align}
\end{subequations}
Using these properties and the dynamics of the empirical average given in Eq.~(\ref{dY}), we find the $n$-adic variation to be
{
    \allowdisplaybreaks
\begin{align}
    &[Y]^{(n)}_t\nonumber\\
    &=\int_{0}^{t} \prod_{k=1}^n\Biggl[\langle \nabla F,\mathbf{K}_{\phi}\phi\rangle_{t'} \mathop{dt'}+D\,\langle \nabla^2 F,\phi\rangle_{t'} \mathop{dt'}+\langle F,\mathbf{H}_\phi\,\phi\rangle_{t'} \mathop{dt'} \nonumber\\
    &\qquad\qquad+\frac{1}{N}\sum_{i_k=1}^N\Big(
        \sum_{j_k=1}^d\partial_{X_{i_k}^{(j_k)}}F_{\mathcal{Y}_{i_k}(t')}(X_{i_k}(t'))\sqrt{2D}\mathop{dW^{(j_k)}_{i_k}(t')}\nonumber\\
     &\qquad\qquad\qquad\qquad+\sum_{\alpha_k\in\mathcal{A}\setminus \{\mathcal{Y}_{i_k}(t')\}}(F_{\alpha_k}(X_{i_k}(t'))-F_{\mathcal{Y}_{i_k}(t')}(X_{i_k}(t')))\mathop{d\hat{N}^{(i_k)}_{\mathcal{Y}_{i_k}(t'),\alpha_k}(t')}\Big)\Biggl]\nonumber\\
    &=\begin{cases}
        Y_t, & n=1,\\
        \frac{1}{N^2}\sum_{i=1}^N  \int_{0}^{t}\Big[2D\sum_{j=1}^d(\partial_{X_i^{(j)}}F_{\mathcal{Y}_i(t')}(X_i(t')))^2\,dt'&\\
        \qquad\qquad\qquad+  \sum_{\alpha\in\mathcal{A}\setminus \{\mathcal{Y}_i(t')\}}(F_\alpha(X_i(t'))-F_{\mathcal{Y}_i(t')}(X_i(t')))^2 d{N}^{(i)}_{\mathcal{Y}_i(t'),\alpha}(t') \Big], & n=2,\\
        \frac{1}{N^n}\sum_{i=1}^N  \int_{0}^{t}\left[\sum_{\alpha\in\mathcal{A}\setminus \{\mathcal{Y}_i(t')\}}(F_\alpha(X_i(t'))-F_{\mathcal{Y}_i(t')}(X_i(t')))^n\mathop{d{N}^{(i)}_{\mathcal{Y}_i(t'),\alpha}(t')} \right], & n>2,
    \end{cases}
    \nonumber\\
    &=\begin{cases}
        Y_t, & n=1,\\
        \frac{2D}{N}\int_0^t\langle (\nabla F)^T\circ\nabla F,\phi\rangle_{t'} dt'&\\
            \quad+\frac{1}{N^2}\sum_{i=1}^N\int_{0}^{t} \Big[\langle F\circ F ,d\mathbf{{N}}_{t'}^{(i)}\phi_i\rangle &\\\qquad\qquad\qquad\qquad\qquad+\langle \mathbf{1} ,d\mathbf{{N}}_{t'}^{(i)}F\circ F\circ \phi_i\rangle -2\langle F ,d\mathbf{{N}}^{(i)}_{t'}F\circ \phi_i\rangle\Big], & n=2,\\
        \frac{1}{N^{n}}\sum_{i=1}^N\int_{0}^{t} \sum_{k=0}^n \binom{k}{n}\langle F^{\circ (n-k)},d\mathbf{{N}}_{t'}^{(i)}(-F)^{\circ  k}\circ \phi_i\rangle_{t'} & n>2,
    \end{cases}
\end{align}
}
where $\circ$ is the Hadamard, or element-wise, product, applied to vectors in $V_\mathcal{A}$, such that, for example, $\langle F^{\circ 2}|= \langle F\circ F|= \langle F|\circ\langle F|= \sum_{\alpha\in\mathcal{A}}\langle \alpha|F^2_\alpha(x)$, and $\langle F^{\circ 0}|\equiv\langle\mathbf{1}|$. 
Note, the binomial sum appearing in the $n>2$ case has simply been explicitly expanded in the $n=2$ case.
\par
Trivially, the expression for $n=1$ is merely $[Y]_t^{(1)}\equiv Y_t$, whilst $n=2$ captures the quadratic variation and thus contains the variance emerging from the Wiener processes. The compensated Poisson noise, however, contributes at all $n>1$ and cannot be written in terms of the total $|\phi\rangle$ due to the presence of the individual particle fields and Poisson increments ($|\phi_i\rangle$ and $dN^{(i)}_{\alpha\beta}$).
\par
However, by taking expectations, through $\mathbb{E}[d\mathbf{N}_t^{(i)}|\phi_i\rangle]=\mathbb{E}[\mathbf{H}_t^{(i)}\,|\phi_i\rangle]\mathop{dt}$, recalling $N^{-1}\sum_{i=1}^N|\phi_i\rangle=|\phi\rangle$, such that $N^{-1}\sum_{i=1}^N\mathbb{E}[d\mathbf{N}_t^{(i)}|\phi_i\rangle]=\mathbb{E}[\mathbf{H}_\phi|\phi\rangle]$, and noting the exchange of order of inner products and expectations by Fubini's theorem (used throughout the subsequent development), we have
\begin{align}
    \mathbb{E}[[Y]^{(n)}_t]&=
    \begin{cases}
        \int_0^t dt'\;\left[\langle \nabla F,\mathbb{E}[\mathbf{K}_{\phi}\phi]\rangle_{t'} +D\langle \nabla^2 F,\mathbb{E}[\phi]\rangle_{t'} +\langle F,\mathbb{E}[\mathbf{H}_\phi\,\phi]\rangle_{t'} \right], & n=1,\\
        \frac{2D}{N}\int_0^t dt' \;\langle (\nabla F)^T\circ \nabla F,\mathbb{E}[\phi]\rangle_{t'} &\\
            \quad+\frac{1}{N}\int_0^tdt'\;\Big[\langle F\circ F ,\mathbb{E}[\mathbf{H}_\phi\,\phi]\rangle_{t'} &\\
            \qquad\qquad\qquad+\langle \mathbf{1} ,\mathbb{E}[\mathbf{H}_\phi\,F\circ F\circ\phi]\rangle_{t'} -2\langle F ,\mathbb{E}[\mathbf{H}_\phi\,F\circ \phi]\rangle_{t'}\Big], & n=2,\\
        \frac{1}{N^{n-1}}\int_0^t dt'\;\sum_{k=0}^n \binom{k}{n}\langle F^{\circ (n-k)},\mathbb{E}[\mathbf{H}_\phi\,(-F)^{\circ  k}\circ\phi]\rangle_{t'}, & n>2,
    \end{cases}
\end{align}
leaving an expression now completely in terms of $|\phi\rangle$. Note, in particular, that all but the mean contribution ($n=1$) vanish as $N\to\infty$ (the limit which facilitates the propagation of chaos result), varying as $N^{1-n}$, in accord with the $\phi^\alpha(x)$ being smooth functions resulting from a deterministic PDE in $p^\alpha(x,t)\equiv\mathbb{E}[\phi^\alpha(x,t)]$. The mean contribution itself is then, naturally, described by Eq.~(\ref{ensembleF}).
\par
As elaborated above, we then consider an alternative dynamic in $\Phi$ (distinct from $\phi$), by replacing the individual noise terms with random fields.
 As with established approaches \cite{deanLangevinEquationDensity1996,worsfoldDensityFluctuationsStochastic2023, tailleurStatisticalMechanicsInteracting2008,chavanisHamiltonianBrownianSystems2008,marconiDynamicDensityFunctional1999,archerDynamicalDensityFunctional2004} we first use  $d+1$ dimensional, vector valued, Brownian sheets, $\mathcal{W}_\alpha (\cdot,\cdot) :\Omega\times\mathbb{R}\to\mathbb{R}^d$, to represent the atomic Wiener processes, but here we require $m$ such objects --- one for each field --- denoted $\mathcal{W}_\alpha(x,t)=(\mathcal{W}^{(1)}_\alpha(x,t),\ldots,\mathcal{W}^{(d)}_\alpha(x,t))$, to replace the individual Wiener processes. Assuming convex $\Omega\ni x,0$, for simplicity, the fields have properties 
 \begin{subequations}
    \begin{align}
    \mathbb{E}[\mathcal{W}^{(i)}_\alpha(x,t)]&=0,\qquad\forall\, i\in\{1,\ldots,d\},\;\forall\,\alpha\in\mathcal{A},\\
    \mathbb{E}[\mathcal{W}^{(i)}_\alpha(x,t)\mathcal{W}^{(j)}_\beta(y,s)]&=\delta_{i,j}\delta_{\alpha,\beta}\,{\rm min}(s,t)\cdot\prod_{k=1}^d{\rm min}(x^{(k)},y^{(k)})\nonumber\\
    &\qquad\forall\, i,j\in\{1,\ldots,d\},\;\forall\,\alpha,\beta\in\mathcal{A},\\
    \int_{0}^x\int_{0}^{t} d\mathcal{W}_\alpha(x',t')&=\mathcal{W}_\alpha(x,t)-\mathcal{W}_\alpha(0,0),\quad\forall\,\alpha\in\mathcal{A},
    \end{align}
 \end{subequations}
 alongside an analogue of the It\^o isometry for expectations of iterated integrals over space 
 \begin{align}
    &\int_{0}^{t}\mathbb{E}\left[\int_{x_1\in\Omega}\int_{x_2\in\Omega}f(x_1)f(x_2)\mathop{d\mathcal{W}^{(i)}_{\alpha}(x_1,t')}\mathop{d\mathcal{W}^{(j)}_{\beta}(x_2,t')}\right]\nonumber\\
    &\qquad\qquad=\delta_{i,j}\delta_{\alpha,\beta}\int_0^tdt'\;\mathbb{E}\left[\int_{\Omega}dx\;f^2(x)\right],\quad\forall\, i,j\in\{1,\ldots,d\},\,\alpha,\beta\in\mathcal{A}.
 \end{align}
 Higher $n$th-order products of $\int_{x\in \Omega} f(x)\mathop{d\mathcal{W}^{(i)}_{\alpha}(x,t)}$ lead to $n$-fold integrals over space with respect to $(d\mathcal{W}^{(i)}_{\alpha}(x,t))^n$, but don't contribute at $\mathcal{O}(dt)$, 
 reflecting the heuristic, generalised, box calculus $d\mathcal{W}^{(i)}_\alpha(x_1^{(i)},t)d\mathcal{W}^{(i)}_\alpha(x_2^{(i)},s)\sim \delta(x_1^{(i)}-x_2^{(i)})\delta(t-s)\mathop{dx}\mathop{dt}$, and the correlative structure of space-time white noise, with the expectation $\mathop{dx}\mathop{dt}$ contributing singularly along the surviving 1D line integral on $x_1=x_2$ (again, to $\mathcal{O}(dt)$). 
 \par
 However, we then diverge from previous approaches by introducing $m\mathop{ (m-1)}$ (one for each possible transition between elements in $\mathcal{A}$) scaled and compensated, Poisson random fields, $\hat{\mathcal{N}}^{N\Phi^\alpha }_{\alpha,\beta}(x,t)$, with increments $d\hat{\mathcal{N}}^{N\Phi^\alpha}_{\alpha,\beta}(x,t)\coloneq d\mathcal{N}_{\alpha,\beta}^{N\Phi^\alpha}(x,t)-\mathbb{H}_{\alpha,\beta}^{N\Phi^\alpha}(x,t)\mathop{dx}\mathop{dt}$, written in terms of an uncompensated version and $\mathbb{H}_{\alpha,\beta}^{N\Phi^\alpha}(x,t)$, which is the spatio-temporal intensity of the fields. We choose such an intensity to be given by 
 \begin{align}
    \mathbb{H}_{\alpha,\beta}^{N\Phi^\alpha}(x,t)\coloneq Nh_{\alpha,\beta}(x,t)\Phi^\alpha(x,t),
    \label{spacetimerate}
 \end{align}
 recalling
 \begin{align}
    h_{\alpha,\beta}(x,t)=  H_{\alpha,\beta}[\langle \mathcal{H}^{\alpha,\beta}_\alpha,\Phi\rangle(x,t)].
 \end{align}
The increments of the uncompensated version are those of an inhomogeneous $d{+}1$ parameter (\emph{i.e.} $d{+}1$ dimensional) Poisson process $d\mathcal{N}^{N\Phi^\alpha}_{\alpha,\beta}(\cdot,\cdot):\Omega\times\mathbb{R}\to\{0,1\}$, which shares the spatio-temporal, field dependent, total intensity, $\mathbb{H}_{\alpha,\beta}^{N\Phi^\alpha}(x,t)$. The compensated field has similar properties to the $\mathcal{W}_\alpha$, again assuming convex $\Omega$ for simplicity,
 \begin{subequations}
 \begin{align}
    \mathbb{E}[\hat{\mathcal{N}}^{N\Phi^\alpha}_{\alpha,\beta}(x,t)]&=0,\quad\forall\,\alpha,\beta\in\mathcal{A},\\
    \int_{0}^{x}\int_{0}^{t} d\hat{\mathcal{N}}^{N\Phi^\alpha}_{\alpha,\beta}(x',t')&=\hat{\mathcal{N}}^{N\Phi^\alpha}_{\alpha,\beta}(x,t)-\hat{\mathcal{N}}^{N\Phi^\alpha}_{\alpha,\beta}(0,0)\nonumber\\
    &={\mathcal{N}}^{N\Phi^\alpha}_{\alpha,\beta}(x,t)-{\mathcal{N}}^{N\Phi^\alpha}_{\alpha,\beta}(0,0)\nonumber\\
    &\qquad-\int_0^tdt'\int_{0}^xdx'\;\mathbb{H}_{\alpha,\beta}^{N\Phi^\alpha}(x',t'),
    \;\forall\,\alpha,\beta\in\mathcal{A},
 \end{align}
\end{subequations}
 but a more complicated covariance structure. It is instead defined through the probability of an event in the uncompensated field, $\mathcal{N}^{N\Phi^\alpha}_{\alpha,\beta}(x,t)$, in a small space-time ``volume''-element, of size $dx\cdot dt$, being given by $\mathbb{H}_{\alpha,\beta}^{N\Phi^\alpha}(x,t)\mathop{dx}\mathop{dt}$, such that $\mathbb{E}[d\mathcal{N}^{N\Phi^\alpha}_{\alpha,\beta}(x,t)|\Phi^\alpha(x,t)]=\mathbb{H}_{\alpha,\beta}^{N\Phi^\alpha}(x,t)\mathop{dx}\mathop{dt}$ and $\mathbb{E}[d\mathcal{N}^{N\Phi^\alpha}_{\alpha,\beta}(x,t)]=\mathbb{E}[\mathbb{H}_{\alpha,\beta}^{N\Phi^\alpha}(x,t)]\mathop{dx}\mathop{dt}$, and where $dx=\prod_{i=1}^d dx_i$. We can view $h_{\alpha,\beta}$ as a base, temporal ``rate per unit time'' component, as with conventional 1D Poisson processes, whilst we additionally have the multiplying $N\Phi^\alpha(x,t)$ as a spatial ``rate per unit space'' component (indicated by the superscript label on the ${\mathcal{N}}^{N\Phi^\alpha}_{\alpha,\beta}$, $\hat{\mathcal{N}}^{N\Phi^\alpha}_{\alpha,\beta}$, and $\mathbb{H}_{\alpha,\beta}^{N\Phi^\alpha}$). Consequently, if all particles were in internal state $\alpha$ (such that $\int_\Omega dx\;\Phi^\alpha(x,t)=1$), and, for simplicity, microscopic transition rates did not depend on the field such that the $h_{\alpha,\beta}(x,t)=h_{\alpha,\beta}$ were constants, we would expect $N\sum_{\gamma\in\mathcal{A}\setminus\{\alpha\}}h_{\alpha,\gamma}$ transitions per unit time on the whole space $\Omega$; an $\sum_{\gamma\in\mathcal{A}\setminus\{\alpha\}}h_{\alpha,\gamma}$ contribution for each particle --- \emph{i.e.} the single particle escape rate from $\alpha$.
\par
The box calculus for these fields follows that of a normal Poisson process, such that, to $\mathcal{O}(1)$, we have, almost surely, $\prod_{i=1}^n d\hat{\mathcal{N}}^{N\Phi^{\alpha_i}}_{\alpha_i,\beta_i}(x_i,t_i)=\prod_{i=1}^n d{\mathcal{N}}^{N\Phi^{\alpha_i}}_{\alpha_i,\beta_i}(x_i,t_i)=d\mathcal{N}^{N\Phi^{\alpha_1}}_{\alpha_1,\beta_1}(x_1,t_1)\prod_{i=1}^{n-1}\delta(x_i-x_{i+1})\delta(t_i-t_{i+1})\delta_{\alpha_i,\alpha_{i+1}}\delta_{\beta_i,\beta_{i+1}}$. It then follows that there is a spatial $n$-fold analogue of an It\^o isometry, where we assume our test function now satisfies $\int_\Omega dx\; f^n(x)<\infty$,
\begin{align}
    &\int_{0}^{t}\mathbb{E}\left[\int_{\{x_1,\ldots,x_n\}\in\Omega^n}\prod_{i=1}^nf(x_i)\mathop{d\hat{\mathcal{N}}^{N\Phi^{\alpha_i}}_{\alpha_i,\beta_i}(x_i,t')}\right]\nonumber\\
    &\qquad\qquad=\int_{0}^{t}\mathbb{E}\left[\int_{x_1\in\Omega}\left(\prod_{i=1}^{n-1}\delta_{\alpha_{i},\alpha_{i+1}}\delta_{\beta_{i},\beta_{i+1}}\right)f^n(x_1)\mathop{d\mathcal{N}^{N\Phi^{\alpha_1}}_{\alpha_1,\beta_1}(x_1,t')}\right]\nonumber\\
    &\qquad\qquad=\left(\prod_{i=1}^{n-1}\delta_{\alpha_{i},\alpha_{i+1}}\delta_{\beta_{i},\beta_{i+1}}\right)\int_{0}^{t}\int_{x'\in\Omega}f^n(x')\mathbb{E}\left[d\mathcal{N}^{N\Phi^{\alpha_1}}_{\alpha_1,\beta_1}(x',t')\right]\nonumber\\
    &\qquad\qquad=N\left(\prod_{i=1}^{n-1}\delta_{\alpha_{i},\alpha_{i+1}}\delta_{\beta_{i},\beta_{i+1}}\right)\int_0^tdt'\int_{\Omega}dx'\;f^n(x')\mathbb{E}\left[h_{\alpha_1,\beta_1}(x',t')\Phi^{\alpha_1}(x',t')\right]\nonumber\\
    &\qquad\qquad=\left(\prod_{i=1}^{n-1}\delta_{\alpha_{i},\alpha_{i+1}}\delta_{\beta_{i},\beta_{i+1}}\right)\int_0^tdt'\int_{\Omega}dx'\;f^n(x')\mathbb{E}\left[\mathbb{H}_{\alpha_1,\beta_1}^{\Phi^{\alpha_1}}(x',t')\right].
\end{align} 
With the relevant behaviour of the random fields established we then collect them into the linear operators $d\boldsymbol{\mathcal{W}}$, where $\langle \beta|d\boldsymbol{\mathcal{W}}|\alpha\rangle = \delta_{\alpha,\beta}\mathop{d\mathcal{W}_\alpha(x,t)}$, and $d\hat{\boldsymbol{\mathcal{N}}}_{\mathbf{H}}^{N\Phi}$, where $\langle\beta|d\hat{\boldsymbol{\mathcal{N}}}_{\mathbf{H}}^{N\Phi}|\alpha\rangle=(1-\delta_{\alpha,\beta}) \mathop{d\hat{\mathcal{N}}^{N\Phi^{\alpha}}_{\alpha,\beta}(x,t)} \allowbreak-\delta_{\alpha,\beta}\sum_{\gamma\in\mathcal{A}\setminus\{\alpha\}}d\hat{\mathcal{N}}^{N\Phi^{\alpha}}_{\alpha,\gamma}(x,t)$,
respectively. The compensated Poisson operator then, also, has a non-compensated version $\langle\beta|d{\boldsymbol{\mathcal{N}}}_{\mathbf{H}}^{N\Phi}|\alpha\rangle=(1-\delta_{\alpha,\beta}) \mathop{d{\mathcal{N}}^{N\Phi^{\alpha}}_{\alpha,\beta}(x,t)}\allowbreak-\delta_{\alpha,\beta}\sum_{\gamma\in\mathcal{A}\setminus\{\alpha\}}d{\mathcal{N}}^{N\Phi^{\alpha}}_{\alpha,\gamma}(x,t)$. Analogously to the individual particle case, we have $\mathbb{E}[d\boldsymbol{\mathcal{N}}_{\mathbf{H}}^{N\Phi}]=N\sum_{\alpha\in\mathcal{A}}\mathbb{E}[\mathbf{H}_\Phi\langle\alpha|\,\Phi\rangle] \mathop{|\alpha\rangle}\langle \alpha|\mathop{dx}\mathop{dt}$ (\emph{i.e.} the expectation of $\mathbf{H}_\phi\mathop{dx}\mathop{dt}$ multiplying a diagonal operator with $\Phi^\alpha$ on the diagonals), or simply $\mathbb{E}[\langle\beta|d\boldsymbol{\mathcal{N}}_{\mathbf{H}}^{N\Phi}|\alpha\rangle]=\mathbb{H}_{\alpha,\beta}^{N\Phi^\alpha}$. We note here that the random fields act as integrators for both space and time, so we clarify the notation of the inner products in these cases to mean $\int_0^t \,\langle F,d\boldsymbol{\mathcal{W}}\Phi\rangle_{t'}\equiv  \sum_{\alpha,\beta\in\mathcal{A}}\int_0^t\int_{x\in\Omega} F_\alpha(x)\Phi^{\alpha}(x,t')\langle\beta|d\boldsymbol{\mathcal{W}}|\alpha\rangle (x,t')=\sum_{\alpha\in\mathcal{A}}\int_0^t\int_{x\in\Omega} F_\alpha(x)\Phi^{\alpha}(x,t')d\mathcal{W}_\alpha (x,t')$.
\par
We then propose the following, alternative, SDE for the empirical average of $\langle F|$, 
$Z(t):=\langle F,\Phi\rangle_t$, in terms of $|\Phi\rangle$, $d\boldsymbol{\mathcal{W}}$ and $d\hat{\boldsymbol{\mathcal{N}}}_{\mathbf{H}}^{N\Phi}$:
 \begin{align}
    {Z}(t)-Z(0)&= \langle F,\Phi\rangle_t-\langle F,\Phi\rangle_0\nonumber\\
    &=\int_0^t\left[-\langle F,\nabla\cdot(\mathbf{K}_{\Phi}\,\Phi)\rangle_{t'} +D\langle F,\nabla^2 \Phi\rangle_{t'}+\langle F,\mathbf{H}_\Phi\,\Phi\rangle_{t'}\right] \mathop{dt'}\nonumber\\
    &\qquad\qquad-\sqrt{2D/N}\int_0^t\langle F,\nabla\cdot (d\boldsymbol{\mathcal{W}}\sqrt{\Phi})\rangle_{t'} + N^{-1}\int_0^t\langle F,d\hat{\boldsymbol{\mathcal{N}}}_{\mathbf{H}}^{N\Phi}\,\mathbf{1}\rangle_{t'},
    \label{altSDE}
 \end{align}
 where $|\sqrt{\Phi}\rangle\equiv|\Phi^{\circ\frac{1}{2}}\rangle=\sum_{\alpha\in\mathcal{A}}\sqrt{\Phi^\alpha(x,t)}|\alpha\rangle$. 
 We then seek to confirm that this SDE replicates the behaviour of that for $Y(t)$ (Eq.~(\ref{dY})), by confirming that is has the same $n$-adic variation, in expectation. To compute the expected $n$-adic variation, $\mathbb{E}[[Z]_t^{(n)}]$, and unlike before, we need to compute averages of $n$-fold iterated integrals and so require the properties above to establish

\begin{align}
    &\mathbb{E}\left[\int_{0}^{t} \langle \nabla F,d\boldsymbol{\mathcal{W}}\sqrt{\Phi}\rangle_{t'}^n\right]\nonumber\\
    &\qquad\mathbb{E}\left[\int_{0}^{t}\prod_{i=1}^n\left(\sum_{\alpha_i\in\mathcal{A}}\int_{x_i\in\Omega}\sum_{j_i=1}^d\left(\partial_{x_i^{(j_i)}}F_{\alpha_i}(x_i)\right)\sqrt{\Phi^{\alpha_i}(x_i,t')}\mathop{d\mathcal{W}^{(j_i)}_{\alpha_i}(x_i,t')}\right)\right]\nonumber\\
    &\qquad=\begin{cases}
        \sum_{\alpha\in\mathcal{A}}\int_0^tdt'\int_\Omega dx \;\left(\nabla_{\!x} F_\alpha(x)\cdot\nabla_{\!x} F_\alpha(x)\right)\mathbb{E}\left[\Phi^\alpha(x,t')\right], & n=2,\\
        0, & n=1\text{ and } n>2,
    \end{cases}
\end{align}
and
\begin{align}
    &\mathbb{E}\left[\int_{0}^{t}\langle F,d\hat{\boldsymbol{\mathcal{N}}}^{N\Phi}_{\mathbf{H}}\mathbf{1}\rangle^n\right]\nonumber\\
        &=\mathbb{E}\left[\int_{0}^{t}\prod_{i=1}^n\left(\sum_{\beta_i\in\mathcal{A}\setminus\{\alpha_i\}}\sum_{\alpha_i\in\mathcal{A}}\int_{x_i\in\Omega}(F_{\beta_i}(x_i)-F_{\alpha_i}(x_i))\mathop{d\hat{\mathcal{N}}^{N\Phi^{\alpha_i}}_{\alpha_i,\beta_i}(x_i,t')}\right)\right]\nonumber\\
        &=\begin{cases}
            0,&n=1,\\
            \sum_{\beta\in\mathcal{A}\setminus\{\alpha\}}\sum_{\alpha\in\mathcal{A}}\int_0^tdt'\int_\Omega dx \;(F_\beta(x)-F_\alpha(x))^nN\mathbb{E}\left[h_{\alpha,\beta}(x,t')\Phi^\alpha(x,t')\right], & n>1,
        \end{cases}\nonumber\\
        &=\begin{cases}
            0,&n=1,\\
            \sum_{\beta\in\mathcal{A}\setminus\{\alpha\}}\sum_{\alpha\in\mathcal{A}}\int_0^tdt'\int_\Omega dx \;(F_\beta(x)-F_\alpha(x))^n\mathbb{E}\left[\mathbb{H}_{\alpha,\beta}^{\Phi^\alpha}(x,t')\right], & n>1.
        \end{cases}
\end{align}
Since we also have $\mathbb{E}[d\mathcal{W}_\alpha(x,t) d\hat{\mathcal{N}}^{N\Phi^{\beta}}_{\beta,\gamma}(y,s)]=0$, due to independence of the random fields, and by counting only $\mathcal{O}(dt)$ contributions in $dZ_t^n$, we can then compute $\mathbb{E}[[Z]^{(n)}_t]$ as
{
\allowdisplaybreaks
\begin{align}
&\mathbb{E}[[Z]^{(n)}_t]\nonumber\\
&=\int_{0}^{t} \mathbb{E}\Bigg[\prod_{i=1}^n\Bigg\{dt'\;\left[\langle \nabla F,\mathbf{K}_{\Phi}\Phi\rangle_{t'} +D\langle \nabla^2 F,\Phi\rangle_{t'} +\langle F,\mathbf{H}_\Phi\,\Phi\rangle_{t'} \right]\nonumber\\
&\qquad\qquad\qquad\qquad+\sqrt{\frac{2D}{N}}\sum_{\alpha_i\in\mathcal{A}}\int_{x_i\in\Omega}\sum_{j_i=1}^d  \left(\partial_{x^{(j_i)}}F_{\alpha_i}(x_i)\right)\sqrt{\Phi^{\alpha_i}(x_i,t')}\mathop{d\mathcal{W}^{(j_i)}_{\alpha_i}(x_i,t')}\nonumber\\
&\qquad\qquad\qquad\qquad+\frac{1}{N}\sum_{\beta_i\in\mathcal{A}\setminus\{\alpha_i\}}\sum_{\alpha_i\in\mathcal{A}}\int_{x_i\in\Omega} (F_{\beta_i}(x_i)-F_{\alpha_i}(x_i))\mathop{d\hat{\mathcal{N}}^{N\Phi^{\alpha_i}}_{\alpha_i,\beta_i}(x_i,t')}\Bigg\}\Bigg]\nonumber\\
    &=\begin{cases}
        \int_0^t dt'\; \mathbb{E}\left[\langle \nabla F,\mathbb{E}[\mathbf{K}_{\Phi}\Phi]\rangle_{t'} +D\langle \nabla^2 F,\mathbb{E}[\Phi]\rangle_{t'} +\langle F,\mathbb{E}[\mathbf{H}_\Phi\Phi]\rangle_{t'} \right],&n=1,\\
        \frac{2D}{N}\int_0^tdt'\;\sum_{\alpha\in\mathcal{A}}\int_\Omega dx\;\left(\nabla_{\!x} F_\alpha(x)\cdot \nabla_{\!x} F_\alpha(x)\right)^2\Phi^\alpha(x,t') dt'&\\
        \quad+\frac{1}{N}\int_0^tdt'\sum_{\alpha\in\mathcal{A}}\sum_{\beta\in\mathcal{A}\setminus\{\alpha\}}\int_\Omega dx\;(F_\beta(x)-F_\alpha(x))^2\mathbb{E}[h_{\alpha,\beta}(x,t')\Phi^\alpha(x,t')] , & n=2,\\
        \frac{1}{N^{n-1}}\int_0^tdt'\sum_{\alpha\in\mathcal{A}}\sum_{\beta\in\mathcal{A}\setminus\{\alpha\}}\int_\Omega dx\;(F_\beta(x)-F_\alpha(x))^n\mathbb{E}[h_{\alpha,\beta}(x,t')\Phi^{\alpha}(x,t')] , & n>2,
        \end{cases}\nonumber\\
        &=\begin{cases}
            \int_0^t dt'\;\left[\langle \nabla F,\mathbb{E}[\mathbf{K}_{\Phi}\Phi]\rangle_{t'} +D\langle \nabla^2 F,\mathbb{E}[\Phi]\rangle_{t'} +\langle F,\mathbb{E}[\mathbf{H}_\Phi\Phi]\rangle_{t'} \right], & n=1,\\
            \frac{2D}{N}\int_0^t dt' \langle (\nabla F)^T\circ\nabla F,\mathbb{E}[\Phi]\rangle_{t'} &\\
                \quad+\frac{1}{N}\int_0^tdt'\;\left[\langle F\circ F ,\mathbb{E}[\mathbf{H}_\Phi\Phi]\rangle_{t'} \right.\\
                \qquad\qquad\qquad\left.+\langle \mathbf{1} ,\mathbb{E}[\mathbf{H}_\Phi F\circ F\circ\Phi]\rangle_{t'} -2\langle F ,\mathbb{E}[\mathbf{H}_\Phi F\circ \Phi]\rangle_{t'}\right], & n=2,\\
            \frac{1}{N^{n-1}}\int_0^t dt'\;\sum_{k=0}^n \binom{k}{n}\langle F^{\circ (n-k)},\mathbb{E}[\mathbf{H}_\Phi(-F)^{\circ  k}\circ\Phi]\rangle_{t'}, & n>2,
        \end{cases}
\end{align}
}
confirming $\mathbb{E}[[Z]_t^{(n)}]=\mathbb{E}[[Y]_t^{(n)}]$, in terms of the field, as required, thus implying the equality in distribution $Z(t)\stackrel{d}{=}Y(t)$. Here we have integrated by parts over the $d\boldsymbol{\mathcal{W}}$ on the first line, and then subsequently enumerated all terms, expanding out all powers of inner products as iterated integrals, kept only those which contribute to $\mathcal{O}(dt)$ under expectations, and then recombined the expression in terms of the total field.
\par
If we then consider Eq.~(\ref{altSDE}) to hold for any suitably well-behaved $\langle F|$, then it implies that the evolution of the fields, $\Phi^\alpha(x,t)$, are described by the non-linear jump-stochastic-integro-differential-equation (implicitly in weak form):
\begin{align}
    \partial_t|\Phi\rangle&=-\nabla\cdot\mathbf{K}_{\Phi}|\Phi\rangle +D\nabla^2|\Phi\rangle +\mathbf{H}_\Phi|\Phi\rangle -\sqrt{\frac{2D}{N}}\nabla\cdot\dot{\boldsymbol{\mathcal{W}}}|\sqrt{\Phi}\rangle +\frac{1}{{N}}\dot{\hat{\boldsymbol{\mathcal{N}}}}^{N\Phi}_{\mathbf{H}}|\mathbf{1}\rangle.
    \label{JSPIDE}
\end{align}
Or, explicitly, without the condensed vector notation, 
\begin{align}
    \label{JSPIDEexplicit}
        \partial_t \Phi^\alpha(x,t)&=-\sum_{\gamma\in\mathcal{A}}\int_\Omega dy\; \nabla_{\!x}\cdot \mathcal{K}_{\alpha,\gamma}(x,y){\Phi}^{\alpha}(x,t){\Phi}^{\gamma}(y,t)+D\nabla^2_{\!x} {\Phi}^{\alpha}(x,t)\nonumber\\
        &\qquad+\sum_{\beta\in\mathcal{A}\setminus\{\alpha\}}H_{\beta,\alpha}\left[\sum_{\gamma\in\mathcal{A}}\int_\Omega dy\;\mathcal{H}^{\beta,\alpha}_{\beta,\gamma}(x,y)\Phi^{\gamma}(y,t)\right]\Phi^{\beta}(x,t)\nonumber\\
        &\qquad-\sum_{\beta\in\mathcal{A}\setminus\{\alpha\}}H_{\alpha,\beta}\left[\sum_{\gamma\in\mathcal{A}}\int_\Omega dy\;\mathcal{H}^{\alpha,\beta}_{\alpha,\gamma}(x,y)\Phi^{\gamma}(y,t)\right]\Phi^{\alpha}(x,t)\nonumber\\
        &\qquad-\sqrt{\frac{2D}{N}}\nabla_{\!x}\cdot\left(\sqrt{\Phi^\alpha(x,t)}\dot{\mathcal{W}}_\alpha(x,t)\right)\nonumber\\
        &\qquad+\frac{1}{N}\sum_{\beta\in\mathcal{A}\setminus\{\alpha\}}\left(\dot{\hat{{\mathcal{N}}}}_{\beta,\alpha}^{N\Phi^{\beta}}(x,t)-\dot{\hat{{\mathcal{N}}}}_{\alpha,\beta}^{N\Phi^{\alpha}}(x,t)\right),\quad\forall\alpha\in\mathcal{A}.
\end{align}
Here the dot notation on the random fields is an abuse of notation indicating distributional sense space-time derivatives of the fields (\emph{i.e.} $\dot{\mathcal{W}}\sim \partial_x\partial_t\mathcal{W}(x,t)$). In the case of the $\dot{\mathcal{W}}_\alpha$ they can be taken as zero mean, Gaussian, space-time white noises. In contrast, the $\dot{\hat{{\mathcal{N}}}}_{\alpha,\beta}^{N\Phi^{\alpha}}$ are manifestly not white due to the smooth compensation terms, but  despite pure Poisson processes readily interpretable as (non-zero mean) white noises, indeed neither are the $\dot{{\mathcal{N}}}_{\alpha\beta}^{N\Phi^{\alpha}}\coloneq \dot{\hat{{\mathcal{N}}}}_{\alpha,\beta}^{N\Phi^{\alpha}}+\mathbb{H}_{\alpha,\beta}^{\Phi^\alpha}$, due to their field dependent inhomogeneity. Regardless, this form of the dynamics, taken in the limit $N\to\infty$, demonstrably leads to a deterministic dynamic in $|\Phi\rangle$, consistent with the propagation of chaos result, as desired.
\par 
Such an equation is then considered to be a generalisation of the DK equation \cite{deanLangevinEquationDensity1996,kawasakiMicroscopicAnalysesDynamical1998}, capturing the (statistics of the) behaviour of the empirical fields formed from It\^o diffusions and Poisson transitions, ostensibly without approximation. We note the incorporation of weak interactions in the sense of McKean \cite{mckeanClassMarkovProcesses1966,jabinMeanFieldLimit2017,chaintronPropagationChaosReview2022, chaintronPropagationChaosReview2022a,worsfoldDensityFluctuationsStochastic2023}, yielding a distinction with DK, which we utilise precisely to exploit the deterministic, $N\to\infty$, limit (\emph{c.f.} Eq.~(\ref{FPMS})). However, a dynamic in terms of strong (DK) interactions follows with appropriately scaled noise terms, see Sec.~\ref{weakvstrong}. We also note, up to such a scaling, the equation reduces to that found in \cite{bressloffGlobalDensityEquations2024} when one averages over the Poisson noise, agreeing at the mean switching level offered in that work.
\par
As with the conventional derivations of DK equations, the above approach is not intended to be rigorous, with the treatments of the random fields almost certainly containing technical deficiencies. Rather, it is intended to heuristically capture the statistical behaviour of the particle noise, including the distinct statistics of the Poisson interconversion events between species, which inevitably contribute at all moments due to their lack of continuity. 
Indeed, the exact nature of Eq.~(\ref{JSPIDE}) itself warrants additional comment. As pointed out in \cite{worsfoldDensityFluctuationsStochastic2023}, once the individual stochastic terms are replaced with the random fields, we have demonstrated, at most, a weak equivalence, or equivalence in distribution/law between $\langle F,\phi\rangle$ and $\langle F,\Phi\rangle$, rather than any stronger notion of equality, \emph{i.e.} almost surely, or in probability etc., nor have we established the nature of the $|\Phi\rangle$ which would satisfy the highly singular Eq.~(\ref{JSPIDE}).
 Moreover, even in the case that $|\mathcal{A}|=m=1$ (such that there is only one density field and no fluctuating Poisson fields), and $\langle \mathcal{K},\Phi\rangle(x)=-(\nabla U*\Phi)(x)$, such that it reduces to the original DK equation (up to a prefactor in $N$), the mathematical status of the equation is rather fraught ---  
it has recently been shown that, the (weak) solutions that do exist correspond only to atomic densities \cite{konarovskyiDeanKawasakiDynamics2020, konarovskyiDeanKawasakiDynamicsIllposedness2019}, leaving the meaning of such equations unclear. Instead, progress has been made in understanding the nature of these equations through regularisation processes, either through the introduction of non-atomic initial data and coloured noise \cite{cornalbaRegularizedDeanKawasaki2019,cornalbaWellposednessRegularisedInertial2021}, or through a cut-off enforced using a spatial discretisation \cite{cornalbaDeanKawasakiEquation2023}. 
However, when the $N\to\infty$ limit is fully saturated, such that the stochastic terms vanish, owing to the weak coupling, the solutions are well-defined in terms of Mckean-Vlasvov diffusions, owing to the propagation of chaos results.
\par
Notably, the formulation in terms of Poisson fields permits a transparent and controlled Gaussian white noise limit as we approach $N\to\infty$, which one can utilise as an approximation in cases where $N\gg 1$. By keeping only leading order contributions to $\mathbb{E}[[Z]_t^{(n)}]$ in $N^{-1}$ we obtain the same quantity if we replace all independent Poisson fields with independent Brownian ones as $d\hat{\mathcal{N}}^{N\Phi^\alpha}_{\alpha,\beta}(x,t)\to \sqrt{\mathbb{H}_{\alpha,\beta}^{N\Phi^\alpha}(x,t)}d\mathcal{W}_{\alpha,\beta}(x,t)$. As such, as $N\to\infty$, we expect the above equation to converge in law to 
\begin{align}
    \partial_t|\Phi\rangle&=-\partial_{x}\mathbf{K}_{\Phi}|\Phi\rangle +D\partial_{x}^2|\Phi\rangle +\mathbf{H}_\Phi|\Phi\rangle -\sqrt{\frac{2D}{N}}\nabla\cdot\dot{\boldsymbol{\mathcal{W}}}|\sqrt{\Phi}\rangle +\frac{1}{\sqrt{N}}\dot{\boldsymbol{\mathcal{W}}}_{\mathbf{H}}|\sqrt{\Phi}\rangle,
    \label{finalB}
\end{align}
where 
\begin{align}
\langle \beta|\dot{\boldsymbol{\mathcal{W}}}_{\mathbf{H}}|\alpha\rangle&=
\begin{cases}
\sqrt{ h_{\alpha,\beta}(x,t)} \dot{\mathcal{W}}_{\alpha,\beta}(x,t),&\alpha\neq \beta,\\
-\sum_{\gamma\in\mathcal{A}\setminus\{ \alpha\}}\sqrt{h_{\alpha,\gamma}(x,t)}\dot{\mathcal{W}}_{\alpha,\gamma}(x,t),&\alpha= \beta,
\end{cases}
\end{align}
and where $\mathcal{W}_{\alpha,\beta}$ are independent Brownian sheets such that $\mathbb{E}[\dot{\mathcal{W}}_{\alpha,\beta}(x,t)\dot{\mathcal{W}}_{\gamma,\delta}(y,s)]=\delta_{\alpha,\gamma}\delta_{\beta,\delta}\delta(x-y)\delta(t-s)$ and $\mathbb{E}[\dot{\mathcal{W}}_{\alpha,\beta}(x,t)\dot{\mathcal{W}}^{(i)}_{\gamma}(y,s)]=0$, for all $\alpha,\beta,\gamma,\delta\in\mathcal{A}$. 
\par
In this large $N$ limit, we are unconcerned with the integration scheme (\emph{e.g.} It\^{o} vs. Stratonovich) associated with these noise terms, but this would require suitable care if extending the approximation beyond this regime. Relatedly, one might wish to more faithfully approximate the higher $N^{-n}$ statistics at smaller $N$, perhaps through an appeal to higher terms in, for example, a van Kampen system size expansion. 
\section{Weak vs. strong interactions}
\label{weakvstrong}
It is important to note that the formulation above is based on weak interactions, with the final SPDE in terms of the empirical one-body \emph{probability} density vector $|\Phi\rangle$. This allows us to rigorously approach a deterministic mean field limit, $\Phi^\alpha(x,t)\to p^\alpha(x,t)$, without recourse to any specific coarse-graining scheme required for a hydrodynamic limit. This property is often useful due to its equivalent (and technically, approximation free) representation as a single particle process in the $N\to\infty$ limit. Generally, however, such a description will most often constitute an approximation to a real interacting system, with it most accurate when the interacting particles are diffuse and the number of effective interactions is large. 
In general, this may well be a tolerable approximation when the interaction kernel represents such a phenomenon or when the interactions are mediated by a coarse-grained mechanism (\emph{e.g.} energetic barrier hopping), as might be the case in the use of the $h_{\alpha\beta}(x,t)$. Similarly, it may be less tolerable when the kernels are representing an intrinsically short-ranged interaction, with behaviour dependent on one, or few, neighbouring particles. However, we can very simply reformulate the resulting SPDE in terms of the physical density field, $|{\varrho}\rangle$, by simply noting that (using the definitions in Dean \cite{deanLangevinEquationDensity1996}) $|\varrho\rangle=N|\Phi\rangle$. Consequently, where a large number of weakly interacting particles on a fixed domain is not desired, but rather a system with strong interactions, we may equivalently express the SPDE as
\begin{align}
    \partial_t|\varrho\rangle&=-\nabla\cdot\mathbf{K}_{\varrho}|\varrho\rangle +D\nabla^2|\varrho\rangle +\mathbf{H}_\varrho|\varrho\rangle -\sqrt{2D}\nabla\cdot\dot{\boldsymbol{\mathcal{W}}}|\sqrt{\varrho}\rangle +\dot{\hat{\boldsymbol{\mathcal{N}}}}^{\varrho}_{\mathbf{H}}|\mathbf{1}\rangle,
\end{align}
with the Poisson fields comprising $\dot{\hat{\boldsymbol{\mathcal{N}}}}^{\varrho}_{\mathbf{H}}$ possessing conditional intensities
\begin{align}
    \mathbb{H}_{\alpha,\beta}^{\varrho^\alpha}(x,t)\coloneq h_{\alpha,\beta}(x,t)\varrho^\alpha(x,t).
 \end{align}
The equivalence, but with a simple scaling in the noise, is the result of the cancellation of the effect of two necessary changes. First, the inner products, $\langle\, \cdot\,,\varrho\rangle$, no longer have an $N^{-1}$ pre-factor when written as a sum, since now  $|\varrho\rangle = \sum_{i=1}^N\delta(x-X_i)|\mathcal{Y}_i\rangle$, such that $\langle \,\cdot\,,\phi\rangle\to N^{-1}\langle \,\cdot\,,\varrho\rangle$. This is then counter-acted by the removal of the weak interaction scaling (the $N^{-1}$ terms) in Eqs.~(\ref{SDEi}) and (\ref{particlerate}), such that $\mathcal{K}\to N \mathcal{K}$ and $\mathcal{H}\to N \mathcal{H}$. Consequently, in aggregate, $\langle \mathcal{K},\phi\rangle\to\langle \mathcal{K},\varrho\rangle$, or $\mathbf{K}_{\phi} \to \mathbf{K}_{\varrho}$, with analogous change $\mathbf{H}_{\phi} \to \mathbf{H}_{\varrho}$. Indeed, this is the conventional formulation of DK type equations. Of course, there is no requirement for all interactions to be of one type, meaning one could stipulate strong interactions mediated by the $\mathcal{K}$, but retain weak interactions mediated by the $\mathcal{H}$.
\par
We note, however, with strong interactions we cannot take the limit $N\to\infty$, which previously lead us to a deterministic total interaction --- due, essentially, to a law of large numbers argument for particles under the interaction kernels --- without causing the physical density and total interaction strength to diverge. Instead, a hydrodynamic limit is obtained in this case through the \emph{joint} limit $N,|\Omega|\to \infty$ such that the total density $N/|\Omega|$ is held constant, subject to a spatial or temporal coarse graining proportional to $|\Omega|$. In such a case, provided the coarse graining occurs above the system's correlation length/time, we would expect to observe that fluctuations in the coarse grained field would vanish as $\sim(\Delta \tau)^{-1/2}$ or $\sim(\Delta x)^{-1/2}$ with $\Delta\tau,\Delta x\propto |\Omega|$ being the time/length scale of the coarse graining. However, in this case the propagation of chaos results no longer hold (as $N,|\Omega|\to\infty$ each particle still locally interacts with a [constant in expectation] finite number of neighbours, meaning the local empirical measure is still stochastic). As a consequence, the resulting hydrodynamic equations will not be closed, with progress reliant on an explicit closure procedure/approximation. The simplest, and perhaps the most common, is merely a mean field assumption which yields precisely the propagation of chaos limit in $|\phi\rangle$, underpinned by weak interactions, thus explicating a consistent microscopic unravelling of this commonly used top-down field-level approximation. This has an implied benefit in applications because it allows one to simulate the underlying stochastic behaviour (SDEs with weak interactions) whilst studying the field level dynamics (\emph{i.e.} through PDEs) that accurately reflect them in a well-defined limit, rather than the more common scenario where underlying fluctuating behaviour is stated and/or simulated (based on strong interactions) which is then approximated with an \emph{ad hoc} mean field assumption at the field level. To be clear, the strongly interacting formulation of the DK equation does accurately reflect the underlying microscopic behaviour, but no limit of that underlying behaviour is accurately reflected, in general, by the mean field assumption applied to the associated DK equation.

\section{Examples}
Following from the above development of the equations of motion for the empirical fields, we now wish to restrict what are highly complicated non-linear dependencies into more manageable, and more plausible, ones through the restriction of the interaction kernels, $\mathcal{K}_{\cdot,\cdot}(\cdot,\cdot)$ and $\mathcal{H}^{\cdot,\cdot}_{\cdot,\cdot}(\cdot,\cdot)$, in order to illustrate the meaning of Eq.~(\ref{JSPIDE}) in some simple examples. In particular, we consider a simple model of run-and-tumble dynamics and one of non-reciprocal flocking behaviour. The purpose of these examples is not to explore the behaviour of the systems, which is well documented elsewhere, but merely to aid understanding of the form of our modified DK equation. We will, however, confirm deterministic ($N\to\infty$) properties of the model with reference to examples in the literature, where appropriate.
\subsection{Example with homogeneous rates: run and tumble active matter}
\label{appRAT}
First, we consider a model of one dimensional run-and-tumble dynamics, capable of motility induced phase separation \cite{catesMotilityInducedPhaseSeparation2015}.
\par
To do so we consider a deterministic interaction kernel of the form
\begin{align}
\mathcal{K}_{\mathcal{Y}_i,\mathcal{Y}_j}(X_i,X_j)=k_0(\mathcal{Y}_i)+k_1(\mathcal{Y}_i)k_\sigma(X_i-X_j),
\end{align}
as applied to the microscopic behaviour, where we stipulate a positive, symmetric, and normalised $k_\sigma(x)$ which takes the limiting from of a Dirac delta measure such that $\lim_{\sigma\to 0}k_\sigma(x)=\delta(x)$ and $\int_\Omega dx\;k_\sigma(x)=1$; for instance $k_\sigma(x)$ could be a normalised top-hat function or Gaussian, with widths $\sigma$. To retain the exact representation, as discussed at the end of Sec.~\ref{DynDenVec}, such kernels would require a point removal at their origin such that $k_\sigma(0)=0$. In the limiting, delta measure, case we can write the pointwise inner products that appear in the dynamics as 
\begin{align}
    \langle \mathcal{K}_{\alpha},\phi\rangle(x)&=\sum_{\beta\in\mathcal{A}}\int_\Omega dy\;\mathcal{K}_{\alpha,\beta}(x,y)\phi^\beta(y)\nonumber\\
&=k_0(\alpha)\sum_{\beta\in\mathcal{A}}\int_{\Omega} dy\; \phi^\beta(y)+k_1(\alpha)\int_{\Omega} dy\; k_\sigma(x-y)\left(\sum_{\beta\in\mathcal{A}}\phi^\beta(y)\right)\nonumber\\
&=k_0(\alpha)\int_{\Omega} dy\; \phi(y)+k_1(\alpha)(k_\sigma *\phi)(x)\nonumber\\
&=k_0(\alpha)+k_1(\alpha)\phi(x)+\mathcal{O}\left(\frac{\sigma^2}{|\Omega|^2}\right),
\end{align}
assuming the field is regular enough on $|\Omega|$ for the convolution to be approximated by a Dirac delta sifting operation, which we expect to be accurate at least in the limit $N\to\infty$ as the field approaches the deterministic propagation of chaos result. Here we have enforced $k_0$, representing intrinsic, non-interacting, behaviour of the particles, to have no dependence on other particles' internal state, $\beta\in\mathcal{A}$, or their own position, $x$, in order to maintain locality and spatial homogeneity, respectively. The component $k_{1}$ then scales a linear particle response to local density with dependence on the particle's internal state.
\par
Note that this limiting procedure, whereby $\sigma/|\Omega|\to 0$, such that the system size is deemed large in comparison to the interaction distance, is distinct from the other limit we have discussed, $N\to\infty$, which is not to do with spatial extent of the system, but to do with the number of weakly coupled particles on a fixed $\Omega$.
\par
To fully construct a model of run-and-tumble interacting behaviour we can specify the following:
\begin{enumerate}
\item We set the domain to be a ring of length $L$, $\Omega=S^1(L/2\pi)$ (equivalent to $\Omega=[-L/2,L/2)$ with periodic boundaries).
\item We consider the set of internal states to be $\mathcal{A}=\{+,-\}$, representing particle orientation on the ring
such that we have
\begin{align}
|\Phi\rangle&=|\Phi^+\rangle+|\Phi^-\rangle=\Phi^+(x,t)|+\rangle+\Phi^-(x,t)|-\rangle.
\end{align}
We then specify the internal state transition rates to be $h_{+,-}(x,t)=h_{+,-}=\gamma$, and $h_{-,+}(x,t)=h_{-,+}=\gamma$ such that the operator $\mathbf{H}_\Phi=\mathbf{H}$ acting on $|\Phi\rangle$ has no field dependence and describes a simple two-level system, represented in terms of a matrix and column vector as
\begin{align}
\mathbf{H}\mathop{|\Phi\rangle}=\begin{bmatrix} -\gamma & \gamma\\ \gamma & -\gamma\end{bmatrix}
\begin{bmatrix}
\Phi^+(x,t)\\
\Phi^-(x,t)
\end{bmatrix}.
\end{align}
\item We choose $k_0(\alpha)=\nu \langle+|\alpha\rangle-\nu \langle-|\alpha\rangle$ where $\nu\geq 0$ is an active motile forcing such that $\alpha$ controls its sign or orientation. Similarly, we choose $k_1(\alpha)=-\nu\xi\langle+|\alpha\rangle+\nu\xi\langle-|\alpha\rangle$, such that increased local density reduces this motile force, in the manner of a quorum sensing interaction \cite{bauerleSelforganizationActiveParticles2018}, where $\xi$ is a coupling strength. We define this coupling strength in terms of the key system parameter $\rho_0\in[0,1]$, which describes the ratio of mean probability density, $\mathbb{E}[\Phi^+(x,t)+\Phi^-(x,t)]=|\Omega|^{-1}=1/L$, to the total empirical probability density, $\Phi^+(x,t)+\Phi^-(x,t)$, which completely arrests the motile forcing under $\mathcal{K}$, such that $\Phi^+(x,t)+\Phi^-(x,t)=\xi^{-1}$. Consequently, we identify $\xi=\rho_0L$. 
\par
The single particle dynamics thus take the form of the SDEs
\begin{subequations}
\begin{align}
X_i(t)&=X_i(0)\nonumber\\
&\quad+\nu\int_0^t dt'\Big[\langle +|\mathcal{Y}_i(t')\rangle-\langle -|\mathcal{Y}_i(t')\rangle\Big]\left(1-\frac{\rho_0L}{N}\sum_{j=1}^Nk_{\sigma}(X_i(t')-X_j(t'))\right) \nonumber\\
&\qquad+\sqrt{2D}\int_0^t\mathop{dW_i}(t'),\\
|\mathcal{Y}_i(t)\rangle&=|\mathcal{Y}_i(0)\rangle+\int_0^t\Big[|+\rangle-|-\rangle\Big]\left[\langle-|\mathcal{Y}_i(t')\rangle \mathop{dN^{(i)}_{-,+}}(t')-\langle+|\mathcal{Y}_i(t')\rangle \mathop{dN^{(i)}_{+,-}}(t')\right],
\end{align} 
\label{finalJSDEs}
\end{subequations}
stipulating the point removal $k_\sigma(0)=0$.

    \par
    Assuming $\sigma\ll L$, this allows us to express the effect of the kernels $\langle\mathcal{K}_\alpha,\Phi\rangle=\sum_{\beta\in\mathcal{A}}\int_\Omega dy\;\mathcal{K}_{\alpha,\beta}(x,y)\phi^\beta(y)$ as $\langle\mathcal{K}_+,\Phi\rangle(x,t)=\nu(1-\rho_0L(\Phi^+(x,t)+\Phi^-(x,t)))$ and $\langle\mathcal{K}_-,\Phi\rangle(x,t)=-\nu(1-\rho_0L(\Phi^+(x,t)+\Phi^-(x,t)))$

\end{enumerate}

Expressing our vectors and operators explicitly as column vectors and matrices for clarity, we may write the resultant form of Eq.~(\ref{JSPIDE}) as
\begin{align}
\partial_t
\begin{bmatrix}
    \Phi^+(x,t)\\
    \Phi^-(x,t)
    \end{bmatrix}
    &=-\partial_x\begin{bmatrix}
        \langle \mathcal{K}_{+},\Phi\rangle(x)&0\\
        0&\langle \mathcal{K}_{-},\Phi\rangle(x)
        \end{bmatrix}
        \begin{bmatrix}
        \Phi^+(x,t)\\
        \Phi^-(x,t)
        \end{bmatrix}
        +D\partial_x^2\begin{bmatrix}
            \Phi^+(x,t)\\
            \Phi^-(x,t)
            \end{bmatrix}\nonumber\\
            &\qquad+\begin{bmatrix} -h_{+,-}(x,t) & h_{-,+}(x,t)\\ h_{+,-}(x,t) & -h_{-,+}(x,t)\end{bmatrix}
            \begin{bmatrix}
            \Phi^+(x,t)\\
            \Phi^-(x,t)
            \end{bmatrix}\nonumber\\
            &\quad-\sqrt{\frac{2D}{N}}\partial_x
            \begin{bmatrix}
                \dot{\mathcal{W}}_+(x,t)&0\\
                0&\dot{\mathcal{W}}_-(x,t)
                \end{bmatrix}
                \begin{bmatrix}
                \sqrt{\Phi^+(x,t)}\\
                \sqrt{\Phi^-(x,t)}
                \end{bmatrix}\nonumber\\
        &\quad+\frac{1}{N}\begin{bmatrix}
            -{\dot{\hat{\mathcal{N}}}^{N\Phi^+}_{+,-}(x,t)}&{\dot{\hat{\mathcal{N}}}^{N\Phi^-}_{-,+}(x,t)}\\
            {\dot{\hat{\mathcal{N}}}^{N\Phi^+}_{+,-}(x,t)}&-{\dot{\hat{\mathcal{N}}}^{N\Phi^-}_{-,+}(x,t)}
            \end{bmatrix}
            \begin{bmatrix}
           1\\
            1
            \end{bmatrix}.
\end{align}
We can then insert the above explicit forms of the kernels and operators obtaining, up to $\mathcal{O}(N^{-1})$ and $\mathcal{O}(\sigma^2/L^2)$,
\begin{align}
    \partial_t
    \begin{bmatrix}
        \Phi^+(x,t)\\
        \Phi^-(x,t)
        \end{bmatrix}
        &=-\partial_x\begin{bmatrix}
            \nu(1-\rho_0L(\Phi^++\Phi^-))&0\\
            0&-\nu(1-\rho_0L(\Phi^++\Phi^-))
            \end{bmatrix}
            \begin{bmatrix}
            \Phi^+(x,t)\\
            \Phi^-(x,t)
            \end{bmatrix}\nonumber\\
            &\quad
            +D\partial_x^2\begin{bmatrix}
                \Phi^+(x,t)\\
                \Phi^-(x,t)
                \end{bmatrix}
               +\begin{bmatrix} -\gamma & \gamma\\ \gamma & -\gamma\end{bmatrix}
                \begin{bmatrix}
                \Phi^+(x,t)\\
                \Phi^-(x,t)
                \end{bmatrix}\nonumber\\
                &\quad-\sqrt{\frac{2D}{N}}\partial_x
                \begin{bmatrix}
                    \dot{\mathcal{W}}_+(x,t)&0\\
                    0&\dot{\mathcal{W}}_-(x,t)
                    \end{bmatrix}
                    \begin{bmatrix}
                    \sqrt{\Phi^+(x,t)}\\
                    \sqrt{\Phi^-(x,t)}
                    \end{bmatrix}\nonumber\\
            &\quad+\frac{1}{N}\begin{bmatrix}
                -{\dot{\hat{\mathcal{N}}}^{N\Phi^+}_{+,-}(x,t)}&{\dot{\hat{\mathcal{N}}}^{N\Phi^-}_{-,+}(x,t)}\\
                {\dot{\hat{\mathcal{N}}}^{N\Phi^+}_{+,-}(x,t)}&-{\dot{\hat{\mathcal{N}}}^{N\Phi^-}_{-,+}(x,t)}
                \end{bmatrix}
                \begin{bmatrix}
               1\\
                1
                \end{bmatrix},
    \end{align}
    or simply

\begin{align}
\partial_t\Phi^\pm(x,t)&=\mp\nu\partial_x\left [\Phi^\pm(x,t)[1-\rho_0 L(\Phi^+(x,t)+\Phi^-(x,t))]\right]\nonumber\\
&\qquad+D\partial_x^2 \Phi^\pm(x,t)\pm\gamma[\Phi^-(x,t)-\Phi^+(x,t)]\nonumber\\
&\qquad-\sqrt{\frac{2D}{N}}\partial_x [\sqrt{\Phi^\pm(x,t)}\dot{\mathcal{W}}_\pm(x,t)]\pm\frac{1}{N}[\dot{\hat{\mathcal{N}}}^{N\Phi^-}_{-,+}(x,t)-\dot{\hat{\mathcal{N}}}^{N\Phi^+}_{+,-}(x,t)]
\end{align}
which can be re-written in terms of the total particle (probability) density $\Phi(x,t)\coloneq \Phi^+(x,t)+\Phi^-(x,t)=\langle\mathbf{1}|\Phi\rangle=\langle+|\Phi\rangle+\langle-|\Phi\rangle$ and polarisation (probability) density $\chi(x,t) \coloneq  \Phi^+(x,t)-\Phi^-(x,t)=\langle+|\Phi\rangle-\langle-|\Phi\rangle$, giving
\begin{subequations}
\begin{align}
\partial_t\Phi(x,t)&=-\nu\partial_x \chi(x,t)\left(1-\rho_0 L\Phi(x,t)\right)+D\partial_x^2 \Phi(x,t)\nonumber\\
&\quad+\sqrt{\frac{D}{N}}\partial_x [\sqrt{\Phi(x,t)+\chi(x,t)}\dot{\mathcal{W}}_+(x,t)+\sqrt{\Phi(x,t)-\chi(x,t)}\dot{\mathcal{W}}_-(x,t)], \\
\partial_t\chi(x,t)&=-\nu\partial_x \Phi(x,t)\left(1-\rho_0 L\Phi(x,t)\right)+D\partial_x^2 \chi(x,t)-2\gamma\chi(x,t) \nonumber\\
&\quad+\sqrt{\frac{D}{N}}\partial_x [\sqrt{\Phi(x,t)+\chi(x,t)}\dot{\mathcal{W}}_+(x,t)-\sqrt{\Phi(x,t)-\chi(x,t)}\dot{\mathcal{W}}_-(x,t)] \nonumber\\
&\quad\qquad+\frac{2}{N}[\dot{\hat{\mathcal{N}}}^{\frac{N}{2}(\Phi-\chi)}_{-,+}(x,t)-\dot{\hat{\mathcal{N}}}^{\frac{N}{2}(\Phi+\chi)}_{+,-}(x,t)].
\end{align}
\end{subequations}
This simply amounts to choosing a desired basis for $V_\mathcal{A}$ in accord with its description as an evolving system of vectors/linear algebra.
\par
Further, in this basis we may describe the Gaussian noise terms as being equal in distribution to two correlated spatio-temporal white noises, each acting on one of $\Phi$ and $\chi$, \emph{viz.}
\begin{subequations}
    \begin{align}
    \partial_t\Phi(x,t)&\stackrel{d}{=}-\nu\partial_x \chi(x,t)\left(1-\rho_0 L\Phi(x,t)\right)+D\partial_x^2 \Phi(x,t)\nonumber\\
    &\quad+\sqrt{\frac{2D}{N}}\partial_x \sqrt{\Phi(x,t)}\dot{\mathcal{W}}_\Phi(x,t),\\
    \partial_t\chi(x,t)&\stackrel{d}{=}-\nu\partial_x \Phi(x,t)\left(1-\rho_0 L\Phi(x,t)\right)+D\partial_x^2 \chi(x,t)-2\gamma\chi(x,t)\nonumber\\
    &\quad+\sqrt{\frac{2D}{N}}\partial_x \sqrt{\Phi(x,t)}\dot{\mathcal{W}}_\chi(x,t) +\frac{2}{N}[\dot{\hat{\mathcal{N}}}^{\frac{N}{2}(\Phi-\chi)}_{-,+}(x,t)-\dot{\hat{\mathcal{N}}}^{\frac{N}{2}(\Phi+\chi)}_{+,-}(x,t)],
    \end{align}
\end{subequations}
where
\begin{subequations}
    \begin{align}
    \mathbb{E}[d\mathcal{W}_\Phi(x,t)d\mathcal{W}_{\chi}(y,s)]&=(\chi(x,t)/\Phi(x,t))\delta(x-y)\delta(t-s)\mathop{dx}\mathop{dt},\\
    \mathbb{E}[d\mathcal{W}_\Phi(x,t)d\mathcal{W}_{\Phi}(y,s)]&=\mathbb{E}[d\mathcal{W}_\chi(x,t)d\mathcal{W}_{\chi}(y,s)]=\delta(x-y)\delta(t-s)\mathop{dx}\mathop{dt}.
    \end{align}
    \label{cov}
    \end{subequations}
Finally, we can describe the system in terms of a conventional P\'eclet number $\text{Pe}=\nu/\sqrt{\gamma D}$ and then perform the 
non-dimensionalisation, $x\to x/l$, $t\to t/\tau$, and $\Phi\to\rho=\Phi/\bar{\Phi}$, $\chi\to\psi=\chi/\bar{\Phi}$, where $l=\sqrt{D/\gamma}$, $\tau=1/\gamma$, and $\bar{\Phi}=(\rho_0 L)^{-1}$ are a characteristic length-scale, timescale, and probability density, respectively. Noting the Brownian rescaling property, $\mathcal{W}(x,t)\stackrel{d}{=}(l\tau)^{-1/2}\mathcal{W}(xl,t\tau)$, alongside the requisite scaling in the Poisson noise (with units probability per unit space per unit time), $\dot{\hat{\mathcal{N}}}\to (\gamma/l)\dot{\hat{\mathcal{N}}}$, we arrive at dynamics in terms of the unit-less particle and polarisation  densities (no longer probability densities), $\rho$ and $\psi$, which vary on the scale of $\rho_0$,
\begin{subequations}
    \begin{align}
\partial_t \rho(x,t)&=-\text{Pe}\partial_x \psi(x,t)\left(1-\rho(x,t)\right)+\partial_x^2\rho(x,t)\nonumber\\
&\qquad+\partial_x\sqrt{2(N_L/N)\rho_0\rho(x,t)}\dot{\mathcal{W}}_{\rho}(x,t),\\
\partial_t \psi(x,t)&=-\text{Pe}\partial_x \rho(x,t)\left(1-\rho(x,t)\right)+\partial_x^2\psi(x,t)-2\psi(x,t)\nonumber\\
&\qquad+\partial_x\sqrt{2(N_L/N)\rho_0\rho(x,t)}\dot{\mathcal{W}}_{\psi}(x,t)\nonumber\\
&\qquad\qquad+\frac{2\rho_0N_L}{N}[\dot{\hat{\mathcal{N}}}^{\frac{N}{2}(\rho-\psi)}_{-,+}(x,t)-\dot{\hat{\mathcal{N}}}^{\frac{N}{2}(\rho+\psi)}_{+,-}(x,t)],
\end{align} 
\label{rescaleRnT}
\end{subequations}
where $N_L=L/l$. Such a transform also manifests itself in the total conditional \emph{rate} of the Poisson noise terms through the scaling $\mathbb{H}^{N\Phi^\alpha}_{\alpha,\beta}(x,t)\to l\tau \mathbb{H}^{N\Phi^\alpha}_{\alpha,\beta}(xl,t\tau)$. Consequently, the Poisson fields appearing in Eq.~(\ref{rescaleRnT}) must then possess rescaled spatio-temporal rates 
\begin{align}
    \mathbb{H}^{\frac{N}{2}(\rho\pm\psi)}_{\pm,\mp}&=l\tau \mathbb{H}^{N\Phi^\pm}_{\pm,\mp}\nonumber\\
    &=l\tau N\gamma \frac{\Phi\pm\chi}{2}\nonumber\\
    &=\frac{Nl}{\rho_0L}\frac{\rho\pm\psi}{2}\nonumber\\
    &=\frac{N}{2\rho_0N_L}(\rho\pm\psi),
\end{align}
which are now dimensionless, also.
\par
The mean behaviour and large $N$ limit of this model, matches the mean behaviour and hydrodynamic limit of a very closely related (and strongly interacting) on-lattice exclusion process \cite{agranovExactFluctuatingHydrodynamics2021, kourbane-housseneExactHydrodynamicDescription2018,erignouxHydrodynamicLimitActive2021}, however it does not possess the same noise structure, neither in terms of the strength of the Gaussian white noises due to the off lattice set-up here, but crucially, we present an equation that retains the Poisson statistics of the tumbling behaviour. 
\par
Taking $N\gg 1$ such that we keep only leading order contributions in $N^{-1}$ in the evolution of moments of the functions of the field from the Poisson terms affects an appropriate Gaussian limit. We first write this as
\begin{subequations}
    \begin{align}
        \partial_t \rho(x,t)&\substack{d\\=\\\lim}-\text{Pe}\partial_x \psi(x,t)\left(1-\rho(x,t)\right)+\partial_x^2\rho(x,t)+\partial_x\sqrt{\frac{2N_L\rho_0\rho(x,t)}{N}}\dot{\mathcal{W}}_{\rho}(x,t),\\
        \partial_t \psi(x,t)&\substack{d\\=\\\lim}-\text{Pe}\partial_x \rho(x,t)\left(1-\rho(x,t)\right)+\partial_x^2\psi(x,t)-2\psi(x,t)+\partial_x\sqrt{\frac{2N_L\rho_0\rho(x,t)}{N}}\dot{\mathcal{W}}_{\psi}(x,t)\nonumber\\
        &\;+\sqrt{\frac{2\rho_0 N_L (\rho(x,t){-}\psi(x,t))}{N}}\dot{\mathcal{W}}_{-,+}(x,t)-\sqrt{\frac{2\rho_0 N_L (\rho(x,t){+}\psi(x,t))}{N}}\dot{\mathcal{W}}_{+,-}(x,t),
    \end{align}
    \end{subequations}
in terms of independent unit spatio-temporal white noises, $\dot{\mathcal{W}}_{+,-}(x,t)$ and $\dot{\mathcal{W}}_{-,+}(x,t)$, the independence of which we can exploit to describe the dynamics, up to equality in distribution, in terms of a single such noise term, uncorrelated with both $\dot{\mathcal{W}}_{\rho}$ and $\dot{\mathcal{W}}_{\psi}$, $\dot{\mathcal{W}}_{\leftrightarrow}(x,t)$,
\begin{subequations}
    \begin{align}
        \partial_t \rho(x,t)&\substack{d\\=\\\lim}-\text{Pe}\partial_x \psi(x,t)\left(1-\rho(x,t)\right)+\partial_x^2\rho(x,t)+\partial_x\sqrt{\frac{2N_L\rho_0\rho(x,t)}{N}}\dot{\mathcal{W}}_{\rho}(x,t),\\
        \partial_t \psi(x,t)&\substack{d\\=\\\lim}-\text{Pe}\partial_x \rho(x,t)\left(1-\rho(x,t)\right)+\partial_x^2\psi(x,t)-2\psi(x,t)\nonumber\\
        &\quad\quad+\partial_x\sqrt{\frac{2N_L\rho_0\rho(x,t)}{N}}\dot{\mathcal{W}}_{\psi}(x,t)+\sqrt{\frac{4N_L\rho_0 \rho(x,t)}{N}}\dot{\mathcal{W}}_{\leftrightarrow}(x,t).
    \end{align}
    \end{subequations}
    This now has all terms explicitly dimensionless, and with noise strength inversely proportional to the (square root of the) number of particles, $N$, per unit-less, rescaled, domain size, $N_L=L/l$, in addition to the total density. This is in agreement with the tumbling noise described in \cite{agranovExactFluctuatingHydrodynamics2021} (before non-dimensionalisation), indicating that the Gaussian limit concurs with contemporary fluctuating hydrodynamic approaches in the same limit. Note, however, that the Poisson processes in these treatments are necessarily considered separately to the remaining dynamics, rendering all fluctuating PDEs as approximations to the behaviour. In contrast, our Eqs.~(\ref{rescaleRnT}) capture all statistics of the Poisson processes as part of the total dynamic.

\subsection{Example with field dependent rates: non-reciprocal ballistic particles with diffusion}
We can provide a less trivial example, where, crucially, transition rates between the states in $\mathcal{A}$ depend on the local field by considering a simple system of non-reciprocal agents \cite{fruchartNonreciprocalPhaseTransitions2021a, kreienkampClusteringFlockingRepulsive2022, youNonreciprocityGenericRoute2020} who, subject to diffusion, move ballistically and align themselves with or against particles of different species based on their own species. 
\par
To do so we consider $|\mathcal{A}|=4$, consisting of $\mathcal{A}=\{A_+,A_-,B_+,B_-\}$, such that there are two `types' of particle, $A$ and $B$, which can each be in a state where they move ballistically to the right or the left. Similarly to the previous example, we consider a ring of length $L$ such that $\Omega = S^1(L/2\pi)$.
\par
Ballistic motion is captured through a non-interacting choice of $\mathcal{K}_{\mathcal{Y}_i,\mathcal{Y}_j}(X_i,X_j)$, specifically
\begin{align}
    \mathcal{K}_{\mathcal{Y}_i,\mathcal{Y}_j}(X_i,X_j)&=k_0(\mathcal{Y}_i)=\nu \left[\langle A_+|\mathcal{Y}_i\rangle+\langle B_+|\mathcal{Y}_i\rangle-\langle A_-|\mathcal{Y}_i\rangle-\langle B_-|\mathcal{Y}_i\rangle\right].
\end{align}
We then implement field dependent rates between species using kernels and mediating functions of the form
\begin{align}
   H_{\alpha,\beta}[\mathcal{H}^{\alpha,\beta}_{\alpha,\epsilon}(x,y)]&= h_0(\alpha,\beta)+\gamma \exp[\mathcal{H}^{\alpha,\beta}_{\alpha,\epsilon}(x,y)],\\
   \mathcal{H}^{\alpha,\beta}_{\alpha,\epsilon}(x,y)&=\lim_{\sigma\to 0}h_1(\alpha,\beta,\epsilon)k_\sigma(x-y),
\end{align}
where we introduce the field independent term
\begin{align}
    h_0(\alpha,\beta)&=\begin{cases}
        \gamma_0, &\alpha = A^{\pm},\,\beta = B^{\pm}\,\,\text{or}\,\,\alpha = B^{\pm},\,\beta = A^{\pm},\\
        0, & \text{otherwise},
    \end{cases}
\end{align}
to allow for ergodicity, and interaction kernel
\begin{align}
    h_1(\alpha,\beta,\epsilon)&=
    \begin{cases}
        \beta_{AA}\left[\langle A_+|\epsilon\rangle-\langle A_-|\epsilon\rangle\right]+\beta_{AB}\left[\langle B_+|\epsilon\rangle-\langle B_-|\epsilon\rangle\right],& \alpha=A_-,\,\beta=A_+,\\
        \beta_{AA}\left[\langle A_-|\epsilon\rangle-\langle A_+|\epsilon\rangle\right]+\beta_{AB}\left[\langle B_-|\epsilon\rangle-\langle B_+|\epsilon\rangle\right],& \alpha=A_+,\,\beta=A_-,\\
        \beta_{BB}\left[\langle B_+|\epsilon\rangle-\langle B_-|\epsilon\rangle\right]+\beta_{BA}\left[\langle A_+|\epsilon\rangle-\langle A_-|\epsilon\rangle\right],& \alpha=B_-,\,\beta=B_+,\\
        \beta_{BB}\left[\langle B_-|\epsilon\rangle-\langle B_+|\epsilon\rangle\right]+\beta_{BA}\left[\langle A_-|\epsilon\rangle-\langle A_+|\epsilon\rangle\right],& \alpha=B_+,\,\beta=B_-,\\
        0, & \text{otherwise},
    \end{cases}
\end{align}
    such that
\begin{align}
    h_{\alpha,\beta} &=H_{\alpha\beta}\left[\langle \mathcal{H}^{\alpha\beta}_{\alpha},\phi\rangle\right]\nonumber\\
    &=\begin{cases}
        \gamma_0, &\alpha = A^{\pm},\,\beta = B^{\pm}\,\,\text{or}\,\,\alpha = B^{\pm},\,\beta = A^{\pm},\\
        \gamma e^{\beta_{AA}(\Phi^{A_+}-\Phi^{A_-})+\beta_{AB}(\Phi^{B_+}-\Phi^{B_-})},& \alpha=A_-,\,\beta=A_+,\\
        \gamma e^{\beta_{AA}(\Phi^{A_-}-\Phi^{A_+})+\beta_{AB}(\Phi^{B_-}-\Phi^{B_+})},& \alpha=A_+,\,\beta=A_-,\\
        \gamma e^{\beta_{BB}(\Phi^{B_+}-\Phi^{B_-})+\beta_{BA}(\Phi^{A_+}-\Phi^{A_-})},& \alpha=B_-,\,\beta=B_+,\\
        \gamma e^{\beta_{BB}(\Phi^{B_-}-\Phi^{B_+})+\beta_{BA}(\Phi^{A_-}-\Phi^{A_+})},& \alpha=B_+,\,\beta=B_-,\\
        0, & \text{otherwise}.
    \end{cases}
\end{align}
This causes species $A$ to align/anti-align with species $A$ according to the size and sign of $\beta_{AA}$, for species $A$ to align/anti-align with species $B$ according to the size and sign of $\beta_{AB}$ \emph{etc.}
\par
Defining the species density and polarisation density fields for convenience, $\Phi_A\coloneq \Phi^{A_+}+\Phi^{A_-}$, $\chi_A\coloneq \Phi^{A_+}-\Phi^{A_-}$, $\Phi_B\coloneq \Phi^{B_+}+\Phi^{B_-}$ and $\chi_B\coloneq \Phi^{B_+}+\Phi^{B_-}$, the non-linear transition matrix and state vector on which it operates are given by
\begin{align}
    \mathbf{H}_\Phi&= \begin{bmatrix} 
        \begin{array}{c|c}
        \mathbf{H}_A&\gamma_0\mathbb{I}\\\hline
        \gamma_0\mathbb{I}&\mathbf{H}_B
        \end{array}
    \end{bmatrix},\\
    \mathbf{H}_A&= \begin{bmatrix} 
        -\gamma_0-\gamma e^{\beta_{AA}\chi_A+\beta_{AB}\chi_B}&\gamma e^{-\beta_{AA}\chi_A-\beta_{AB}\chi_B}\\
        \gamma e^{\beta_{AA}\chi_A+\beta_{AB}\chi_B}&-\gamma_0-\gamma e^{-\beta_{AA}\chi_A-\beta_{AB}\chi_B}
    \end{bmatrix},\\
    \mathbf{H}_B&= \begin{bmatrix} 
        -\gamma_0-\gamma e^{\beta_{BB}\chi_B+\beta_{BA}\chi_A}&\gamma e^{-\beta_{BB}\chi_B-\beta_{BA}\chi_A}\\
        \gamma e^{\beta_{BB}\chi_B+\beta_{BA}\chi_A}&-\gamma_0-\gamma e^{-\beta_{BB}\chi_B-\beta_{BA}\chi_A}
    \end{bmatrix},\\
  |\Phi\rangle&=
    \begin{bmatrix}
        \Phi^{A_+}(x,t)\\
        \Phi^{A_-}(x,t)\\
        \Phi^{B_+}(x,t)\\
        \Phi^{B_-}(x,t)
    \end{bmatrix}.
    \end{align}
    Inserting these kernel and mediating functions into Eq.~(\ref{JSPIDE}), we find
    \begin{subequations}
    \begin{align}
        \partial_t\Phi^{A_+}&=-\nu\partial_x\Phi^{A_+}+D\partial_x^2\Phi^{A_+}-\gamma e^{\beta_{AA}\chi_A+\beta_{AB}\chi_B}\Phi^{A_+}+\gamma e^{-\beta_{AA}\chi_A-\beta_{AB}\chi_B}\Phi^{A_-}\nonumber\\
        &\qquad\qquad+\gamma_0(\Phi^{B_+}-\Phi^{A_+})-\sqrt{\frac{2D}{N}}\partial_x\left[\sqrt{\Phi^{A_+}}dW_{A_+}\right]\nonumber\\
        &\qquad\qquad\qquad\qquad+\frac{1}{N}\left[\dot{\hat{\mathcal{N}}}^{N\Phi^{A_-}}_{A_-,A_+}-\dot{\hat{\mathcal{N}}}^{N\Phi^{A_+}}_{A_+,A_-}+\dot{\hat{\mathcal{N}}}^{N\Phi^{B_+}}_{B_+,A_+}-\dot{\hat{\mathcal{N}}}^{N\Phi^{A_+}}_{A_+,B_+}\right],\\
        \partial_t\Phi^{A_-}&=+\nu\partial_x\Phi^{A_+}+D\partial_x^2\Phi^{A_+}+\gamma e^{\beta_{AA}\chi_A+\beta_{AB}\chi_B}\Phi^{A_+}-\gamma e^{-\beta_{AA}\chi_A-\beta_{AB}\chi_B}\Phi^{A_-}\nonumber\\
        &\qquad\qquad+\gamma_0(\Phi^{B_-}-\Phi^{A_-})-\sqrt{\frac{2D}{N}}\partial_x\left[\sqrt{\Phi^{A_-}}dW_{A_-}\right]\nonumber\\
        &\qquad\qquad\qquad\qquad+\frac{1}{N}\left[\dot{\hat{\mathcal{N}}}^{N\Phi^{A_+}}_{A_+,A_-}-\dot{\hat{\mathcal{N}}}^{N\Phi^{A_-}}_{A_-,A_+}+\dot{\hat{\mathcal{N}}}^{N\Phi^{B_-}}_{B_-,A_-}-\dot{\hat{\mathcal{N}}}^{N\Phi^{A_-}}_{A_-,B_-}\right],
    \end{align}
    and
    \begin{align}
        \partial_t\Phi^{B_+}&=-\nu\partial_x\Phi^{B_+}+D\partial_x^2\Phi^{B_+}-\gamma e^{\beta_{BB}\chi_B+\beta_{BA}\chi_A}\Phi^{B_+}+\gamma e^{-\beta_{BB}\chi_B-\beta_{BA}\chi_A}\Phi^{B_-}\nonumber\\
        &\qquad\qquad+\gamma_0(\Phi^{A_+}-\Phi^{B_+})-\sqrt{\frac{2D}{N}}\partial_x\left[\sqrt{\Phi^{B_+}}dW_{B_+}\right]\nonumber\\
        &\qquad\qquad\qquad\qquad+\frac{1}{N}\left[\dot{\hat{\mathcal{N}}}^{N\Phi^{B_-}}_{B_-,B_+}-\dot{\hat{\mathcal{N}}}^{N\Phi^{B_+}}_{B_+,B_-}+\dot{\hat{\mathcal{N}}}^{N\Phi^{A_+}}_{A_+,B_+}-\dot{\hat{\mathcal{N}}}^{N\Phi^{B_+}}_{B_+,A_+}\right],\\
        \partial_t\Phi^{B_-}&=+\nu\partial_x\Phi^{B_+}+D\partial_x^2\Phi^{B_+}+\gamma e^{\beta_{BB}\chi_B+\beta_{BA}\chi_A}\Phi^{B_+}-\gamma e^{-\beta_{BB}\chi_B-\beta_{BA}\chi_A}\Phi^{B_-}\nonumber\\
        &\qquad\qquad+\gamma_0(\Phi^{A_-}-\Phi^{B_-})-\sqrt{\frac{2D}{N}}\partial_x\left[\sqrt{\Phi^{B_-}}dW_{B_-}\right]\nonumber\\
        &\qquad\qquad\qquad+\frac{1}{N}\left[\dot{\hat{\mathcal{N}}}^{N\Phi^{B_+}}_{B_+,B_-}-\dot{\hat{\mathcal{N}}}^{N\Phi^{B_-}}_{B_-,B_+}+\dot{\hat{\mathcal{N}}}^{N\Phi^{A_-}}_{A_-,B_-}-\dot{\hat{\mathcal{N}}}^{N\Phi^{B_-}}_{B_-,A_-}\right].
    \end{align}
\end{subequations}
    We can then write this in terms of the particle type density and polarisation density fields $\Phi_{A/B}$ and $\chi_{A/B}$, in which case we have, along with aggregation of Gaussian noise terms into four correlated terms for each field,
\begin{subequations}
    \begin{align}
        \partial_t\Phi_A&\stackrel{d}{=}-\nu\partial_x\chi_A+D\partial_x^2\Phi_A+\gamma_0(\rho_B-\rho_A)+\sqrt{\frac{2D}{N}}+ \partial_x[\sqrt{\Phi_A}\dot{\mathcal{W}}_{\Phi_A}]\nonumber\\
        &\qquad+\frac{1}{N}\left[\dot{\hat{\mathcal{N}}}^{\frac{N}{2}(\Phi_B+\chi_B)}_{B_+,A_+}+\dot{\hat{\mathcal{N}}}^{\frac{N}{2}(\Phi_B-\chi_B)}_{B_-,A_-}-\dot{\hat{\mathcal{N}}}^{\frac{N}{2}(\Phi_A-\chi_A)}_{A_-,B_-}-\dot{\hat{\mathcal{N}}}^{\frac{N}{2}(\Phi_A+\chi_A)}_{A_+,B_+}\right],\\
        \partial_t\chi_A&\stackrel{d}{=}-\nu\partial_x\Phi_A+D\partial_x^2\chi_A+\gamma_0(\chi_B-\chi_A)+\sqrt{\frac{2D}{N}}\partial_x[\sqrt{\Phi_A}\dot{\mathcal{W}}_{\chi_A}]\nonumber\\
        &\qquad+\Phi_A\sinh[\beta_{AA}\chi_A+\beta_{AB}\chi_B]-\chi_A\cosh[\beta_{AA}\chi_A+\beta_{AB}\chi_B]\nonumber\\
        &\qquad+\frac{2}{N}\left[\dot{\hat{\mathcal{N}}}^{\frac{N}{2}(\Phi_A-\chi_A)}_{A_-,A_+}-\dot{\hat{\mathcal{N}}}^{\frac{N}{2}(\Phi_A+\chi_A)}_{A_+,A_-}+\dot{\hat{\mathcal{N}}}^{\frac{N}{2}(\Phi_B+\chi_B)}_{B_+,A_+}-\dot{\hat{\mathcal{N}}}^{\frac{N}{2}(\Phi_B-\chi_B)}_{B_-,A_-}\right.\nonumber\\
        &\qquad\qquad\qquad\left.+\dot{\hat{\mathcal{N}}}^{\frac{N}{2}(\Phi_A-\chi_A)}_{A_-,B_-}-\dot{\hat{\mathcal{N}}}^{\frac{N}{2}(\Phi_A+\chi_A)}_{A_+,B_+}\right],
    \end{align}
    and
    \begin{align}
        \partial_t\Phi_B&\stackrel{d}{=}-\nu\partial_x\chi_B+D\partial_x^2\Phi_B+\gamma_0(\rho_A-\rho_B)+\sqrt{\frac{2D}{N}}+ \partial_x[\sqrt{\Phi_B}\dot{\mathcal{W}}_{\Phi_B}]\nonumber\\
        &\qquad+\frac{1}{N}\left[\dot{\hat{\mathcal{N}}}^{\frac{N}{2}(\Phi_A+\chi_A)}_{A_+,B_+}+\dot{\hat{\mathcal{N}}}^{\frac{N}{2}(\Phi_A-\chi_A)}_{A_-,B_-}-\dot{\hat{\mathcal{N}}}^{\frac{N}{2}(\Phi_B-\chi_B)}_{B_-,A_-}-\dot{\hat{\mathcal{N}}}^{\frac{N}{2}(\Phi_B+\chi_B)}_{B_+,A_+}\right],\\
        \partial_t\chi_B&\stackrel{d}{=}-\nu\partial_x\Phi_B+D\partial_x^2\chi_B+\gamma_0(\chi_A-\chi_B)+\sqrt{\frac{2D}{N}}\partial_x[\sqrt{\Phi_B}\dot{\mathcal{W}}_{\chi_B}]\nonumber\\
        &\qquad+\Phi_B\sinh[\beta_{BB}\chi_B+\beta_{BA}\chi_A]-\chi_B\cosh[\beta_{BB}\chi_B+\beta_{BA}\chi_A]\nonumber\\
        &\qquad+\frac{2}{N}\left[\dot{\hat{\mathcal{N}}}^{\frac{N}{2}(\Phi_B-\chi_B)}_{B_-,B_+}-\dot{\hat{\mathcal{N}}}^{\frac{N}{2}(\Phi_B+\chi_B)}_{B_+,B_-}+\dot{\hat{\mathcal{N}}}^{\frac{N}{2}(\Phi_A+\chi_A)}_{A_+,B_+}-\dot{\hat{\mathcal{N}}}^{\frac{N}{2}(\Phi_A-\chi_A)}_{A_-,B_-}\right.\nonumber\\
        &\qquad\qquad\qquad\left.+\dot{\hat{\mathcal{N}}}^{\frac{N}{2}(\Phi_B-\chi_B)}_{B_-,A_-}-\dot{\hat{\mathcal{N}}}^{\frac{N}{2}(\Phi_B+\chi_B)}_{B_+,A_+}\right],
    \end{align}
\end{subequations}
where
\begin{subequations}
    \begin{align}
        \mathbb{E}[\dot{\mathcal{W}}_{\Phi_A}(x,t)\dot{\mathcal{W}}{\chi_A}(y,s)]&=(\chi_A(x,t)/\Phi_A(x,t))\delta(x-y)\delta(t-s),\\
        \mathbb{E}[\dot{\mathcal{W}}_{\Phi_B}(x,t)\dot{\mathcal{W}}_{\chi_B}(y,s)]&=(\chi_B(x,t)/\Phi_B(x,t))\delta(x-y)\delta(t-s),\\
    \mathbb{E}[\dot{\mathcal{W}}_{\Phi_A}(x,t)\dot{\mathcal{W}}_{\Phi_B}(y,s)]&=\mathbb{E}[\dot{\mathcal{W}}_{\Phi_A}(x,t)\dot{\mathcal{W}}_{\chi_B}(y,s)]=\mathbb{E}[\dot{\mathcal{W}}_{\chi_A}(x,t)\dot{\mathcal{W}}_{\Phi_B}(y,s)]\nonumber\\
    &=\mathbb{E}[\dot{\mathcal{W}}_{\chi_A}(x,t)\dot{\mathcal{W}}_{\chi_B}(y,s)]=0,\\    
    \mathbb{E}[\dot{\mathcal{W}}_{\Phi_A}(x,t)\dot{\mathcal{W}}_{\Phi_A}(y,s)]&=\mathbb{E}[\dot{\mathcal{W}}_{\Phi_B}(x,t)\dot{\mathcal{W}}_{\Phi_B}(y,s)]=\mathbb{E}[\dot{\mathcal{W}}_{\chi_A}(x,t)\dot{\mathcal{W}}_{\chi_A}(y,s)]\nonumber\\
    &=\mathbb{E}[\dot{\mathcal{W}}_{\chi_B}(x,t)\dot{\mathcal{W}}_{\chi_B}(y,s)]= \delta(x-y)\delta(t-s).
    \end{align}
    \end{subequations}
Here we have refrained from a space or time rescaling of the equations and thus the spatio-temporal Poisson rates are simply given by their definition (c.f. Eq.~(\ref{spacetimerate}))
\begin{align}
    \mathbb{H}^{N\Phi^{\alpha}}_{\alpha,\beta}&=N\Phi^{\alpha}(x,t)h_{\alpha,\beta}(x,t).
\end{align}
When linearised around $\{\Phi_0/2,0,\Phi_0/2,0\}$ and considered in reciprocal space in the $|\Phi'\rangle=\{\Phi_A,\chi_A,\Phi_B,\chi_B\}$ basis, where $\Phi_0=\sum_{\alpha\in\{A,B\}}\int_0^Ldx\,\Phi_\alpha(x,t)$, the deterministic component of the system, $-\nabla\cdot\mathbf{K}_{\Phi'}+D\nabla^2 +\mathbf{H}_{\Phi'}$,  can be compactly described through the matrix
\begin{align} 
    &\mathbf{M}\nonumber\\&=\left(
        \begin{array}{cccc}
         -\gamma _0-D k^2 & -i k \nu  & \gamma _0 & 0 \\
         -i k \nu  & \frac{1}{2} \gamma  \Phi _0 \beta _{AA}-\gamma -\gamma _0-D k^2 & 0 & \frac{1}{2} \gamma  \Phi _0 \beta _{AB}+\gamma _0 \\
         \gamma _0 & 0 & -\gamma _0-D k^2 & -i k \nu  \\
         0 & \frac{1}{2} \gamma  \Phi _0 \beta _{BA}+\gamma _0 & -i k \nu  & \frac{1}{2} \gamma  \Phi _0 \beta _{BB}-\gamma -\gamma _0-D k^2 \\
        \end{array}
        \right),
\end{align}  
describing a set of ODEs for the modes of the system in the $N\to\infty$, propagation of chaos, limit. Linear response can be probed through the largest eigenvalue of this matrix, $\lambda_{\rm max}$, which in the $k\to 0$ limit, relevant on long, hydrodynamic, scales, is given by
\begin{align}
    4\lambda_{\rm max} &=\sqrt{\gamma ^2 \Phi _0^2 \left(\left(\beta _{AA}-\beta _{BB}\right)^2+4 \beta _{AB} \beta _{BA}\right)+8 \gamma 
    \gamma _0 \Phi _0 \left(\beta _{AB}+\beta _{BA}\right)+16 \gamma _0^2}\nonumber\\
    &\qquad\qquad\qquad+\gamma  \left(\Phi _0 \left(\beta _{AA}+\beta
    _{BB}\right)-4\right)-4 \gamma _0.
\end{align}
The system exhibits travelling oscillatory behaviour, characteristic of a chase-and-catch dynamic resulting from a non-reciprocal interaction kernel, when the imaginary component of the eigenvalue becomes non-zero. In the (non-ergodic) limit $\gamma_0\to 0$ this condition is met when
\begin{align}
    (\beta_{BB}-\beta_{AA})^2<-4\beta_{AB}\beta_{BA},
\end{align}
such that interspecies couplings $\beta_{AB}$ and $\beta_{BA}$ have opposite signs (\emph{i.e.} a non-reciprocal interaction), which dominate over self alignment, characteristic of a flocking interaction. For completeness, we note that the spinodals of a flocking transition occur when the real component of the eigenvalue becomes positive in this limit.
\par
This behaviour completely mirrors the results in \cite{kreienkampClusteringFlockingRepulsive2022} which models active chiral systems. Here the 1D nature of the system is what gives rise to the Poisson transitions, rather than continuous rotational alignment dynamics in \cite{kreienkampClusteringFlockingRepulsive2022}, however the limiting deterministic character is retained, here with an exact representation of the Poisson transitions.
\section{Non-conservative, binary and higher order reactions}
The preceding development has presented an equation based solely on unary reactions of the form
\begin{subequations}
\begin{align}
    A\rightarrow B,\\
    B\rightarrow A,
\end{align}
\end{subequations}
albeit with non-linear and spatially varying, density dependent rates.
\par
However, it would be of interest to describe reactions where particle number were not conserved, such as 
\begin{subequations}
\begin{align}
    \emptyset&\rightarrow A,\\
    A&\rightarrow \emptyset,
\end{align}
\end{subequations}
and those of higher order, such as
\begin{subequations}
\begin{align}
    2A&\rightarrow B,\\
    C&\rightarrow A + B,\\
    A+B&\rightarrow C + D,
\end{align}
\end{subequations}
and so forth.
\par
The matter of deriving an SPDE for such processes is left for future work, however, we can describe a possible microscopic set up that would allow such reactions to be possible.
\par
First, we consider the binary reaction
\begin{align}
    A+B&\rightarrow C + D,
\end{align}
which conserves particle number. Here we have a situation we can describe by having a correlated event whereupon a particle of type $A$ transitions to one of type $C$, and a particle of type $B$ transitions to one of type $D$. We can achieve the necessary correlation by considering underlying driving Poisson processes which are shared by each possible unordered pair of particles of type $A$ and $B$. That is, the dynamic in the underlying particle becomes
\begin{align}
    |\mathcal{Y}_i(t)\rangle-|\mathcal{Y}_i(0)\rangle &=\int_0^t\sum_{\alpha\in\mathcal{A}\setminus\{\mathcal{Y}_i(t')\}}\Big[|\alpha\rangle-|\mathcal{Y}_i(t')\rangle\Big]\mathop{dN^{(i)}_{{\mathcal{Y}_i,\alpha}}}(t')\nonumber\\
    &\qquad+\sum_{j=1,j\neq i}^N\int_0^t\delta_{\mathcal{Y}_i(t'),A}\delta_{\mathcal{Y}_j(t'),B}\Big[|C\rangle-|\mathcal{Y}_i(t')\rangle\Big]dN_{AB,CD}^{(i,j)}(t')\nonumber\\
    &\qquad+\sum_{j=1,j\neq i}^N\int_0^t\delta_{\mathcal{Y}_i(t'),B}\delta_{\mathcal{Y}_j(t'),A}\Big[|D\rangle-|\mathcal{Y}_i(t')\rangle\Big]dN_{AB,CD}^{(i,j)}(t'),
\end{align}
for all $i\in\{1,\ldots,N\}$, 
such that a transition in particle $i$ from state $A$ to $C$ is always coincident with a transition in particle $j$ from state $B$ to $D$; \emph{i.e.} we stipulate $dN_{AB,CD}^{(i,j)}\equiv dN_{AB,CD}^{(j,i)}$. The kernel functions that comprise the transition rate would then responsible for determining if the requisite density and locality conditions, for and between the particles, are met to facilitate a transition. Higher order reactions (\emph{e.g.} ternary, quaternary \emph{etc.}) would then be formulated in an analogous manner.
\par
In contrast, reactions that do not conserve particle number need a distinct approach. We postulate that a suitable manner of achieving such reactions is through the use of a null state, $\emptyset$, such that $\mathcal{A}\to\mathcal{A}\cup\{\emptyset\}$, with the null state acting as a reservoir of generic `particles' which can be added to, or drawn from, as required. 
Thus, reactions of type $A+B\to C$ can be reinterpreted as $A+B\to C+\emptyset$, \emph{etc.} 
\par 
A delicate question in such circumstances is how to interpret the particle number in this instance. We propose that the number $N$ should itself be treated as a dynamical variable, requiring updates in the shared Poisson processes. For instance, in a single species model which is spontaneously annihilated and created according to $A\to\emptyset$ and $\emptyset\to A$, one could keep track of the total particle number with a process
\begin{align}
    N(t)-N(0)&=\sum_{i=1}^N\left[\int_0^t dN^{(i)}_{\emptyset, A}(t')-\int_0^tdN^{(i)}_{A,\emptyset}(t')\right],
\end{align}
comprised of Poisson increments responsible for updating the individual particles. We thus anticipate that in addition to the resultant SPDEs there would be a coupled and correlated, jump type, SDE governing $N$ for which suitable approximations could be made (\emph{e.g.} a van Kampen system size expansion). Equivalently, one can imagine that the approach for unary reactions offered in this contribution is implicitly coupled to the trivial SDE, $dN=0$.
\section{Discussion and Conclusion}
In this contribution we have introduced a stochastic partial differential equation (Eq.~(\ref{JSPIDE})), driven by both Brownian and Poisson random fields, which is equal in law with a set of point particles evolving according to first order stochastic differential equations with unary reactions between internal states. This generalises the DK equation to this class of reaction diffusion system, concisely representing the fluctuating density dynamics, without approximation.
\par
We have chosen a limited class of SDEs, and their coupling to the field, essentially for readability, however we note the obvious generalisations that can be made to this scheme. For instance, even the most general interaction used in this work, captured through the Poisson transitions, 
could be generalised so as to introduce distinct mediating functions based on the different density fields so as to allow for non-linear rates which differ significantly based on the species with which the particles are interacting. In contrast, the continuous deterministic interaction, $\mathbf{K}_{\Phi}$, has been presented with no such mediating function, which could naturally be added. Similarly, we have assumed isotropic diffusion and independence in noise strength with both particle state, and the state of the field, all of which are eminently possible under this scheme. For instance, for work concerning noise strength that depends on the field, with the use of a particular mediating function, see \cite{worsfoldDensityFluctuationsStochastic2023}. The augmentation of the dynamics to include such additional complication is not expected to add any particular technical obstacles, however we note that, whilst propagation of chaos limits are generally also expected under such generalisations, the speed at which convergence happens is not guaranteed to be equal, nor to be fast \cite{chaintronPropagationChaosReview2022,chaintronPropagationChaosReview2022a}.
\par
We have presented two simple examples, with their purpose being to illustrate the meaning of Eq.~(\ref{JSPIDE}) in various contexts, rather than for them to be studied deeply. As such we have merely demonstrated that they agree with known models where relevant, and point out that Eq.~(\ref{JSPIDE}) has led to this agreement whilst retaining the exact fluctuation properties of the Poisson jumps. The application of such an equation to produce approximations  beyond that of Gaussian statistics is a logical and pertinent next step for this work, as is the possibility of generating numerical approximations of Eq.~(\ref{JSPIDE}), but are ones we leave for future contributions.

%% file: references.tex
\providecommand{\newblock}{}